\documentclass[journal=aamick,manuscript=article]{achemso}
\usepackage{chemformula} 
\usepackage[T1]{fontenc} 
\usepackage{ulem}

\author{Katarzyna Ludwiczak}
\affiliation[University of Warsaw]
{Faculty of Physics, University of Warsaw, ul. Pasteura 5, 02-093 Warsaw}
\email{kw.ludwiczak@uw.edu.pl}

\author{Aleksandra Krystyna Dąbrowska}
\affiliation[University of Warsaw]
{Faculty of Physics, University of Warsaw, ul. Pasteura 5, 02-093 Warsaw}
\author{Johannes Binder}
\affiliation[University of Warsaw]
{Faculty of Physics, University of Warsaw, ul. Pasteura 5, 02-093 Warsaw}
\author{Mateusz Tokarczyk}
\affiliation[University of Warsaw]
{Faculty of Physics, University of Warsaw, ul. Pasteura 5, 02-093 Warsaw}
\author{Jakub Iwański}
\affiliation[University of Warsaw]
{Faculty of Physics, University of Warsaw, ul. Pasteura 5, 02-093 Warsaw}
\author{Bogusława Kurowska}
\affiliation[Polish Academy of Sciences]
{Institute of Physics Polish Academy of Sciences, Al. Lotników 32/46, 02-668 Warsaw}
\author{Jakub Turczyński}
\affiliation[Polish Academy of Sciences]
{Institute of Physics Polish Academy of Sciences, Al. Lotników 32/46, 02-668 Warsaw}
\author{Grzegorz Kowalski}
\affiliation[University of Warsaw]
{Faculty of Physics, University of Warsaw, ul. Pasteura 5, 02-093 Warsaw}
\author{Rafał Bożek}
\affiliation[University of Warsaw]
{Faculty of Physics, University of Warsaw, ul. Pasteura 5, 02-093 Warsaw}
\author{Roman Stępniewski}
\affiliation[University of Warsaw]
{Faculty of Physics, University of Warsaw, ul. Pasteura 5, 02-093 Warsaw}
\author{Wojciech Pacuski}
\affiliation[University of Warsaw]
{Faculty of Physics, University of Warsaw, ul. Pasteura 5, 02-093 Warsaw}
\author{Andrzej Wysmołek}
\affiliation[University of Warsaw]
{Faculty of Physics, University of Warsaw, ul. Pasteura 5, 02-093 Warsaw}

\title[Heteroepitaxial growth of high optical quality, wafer-scale van der Waals heterostrucutres]
  {Heteroepitaxial growth of high optical quality, wafer-scale van der Waals heterostrucutres}

\abbreviations{TMD, MOVPE, MBE}

\begin{document}

\begin{tocentry}
\includegraphics[width=1\textwidth]{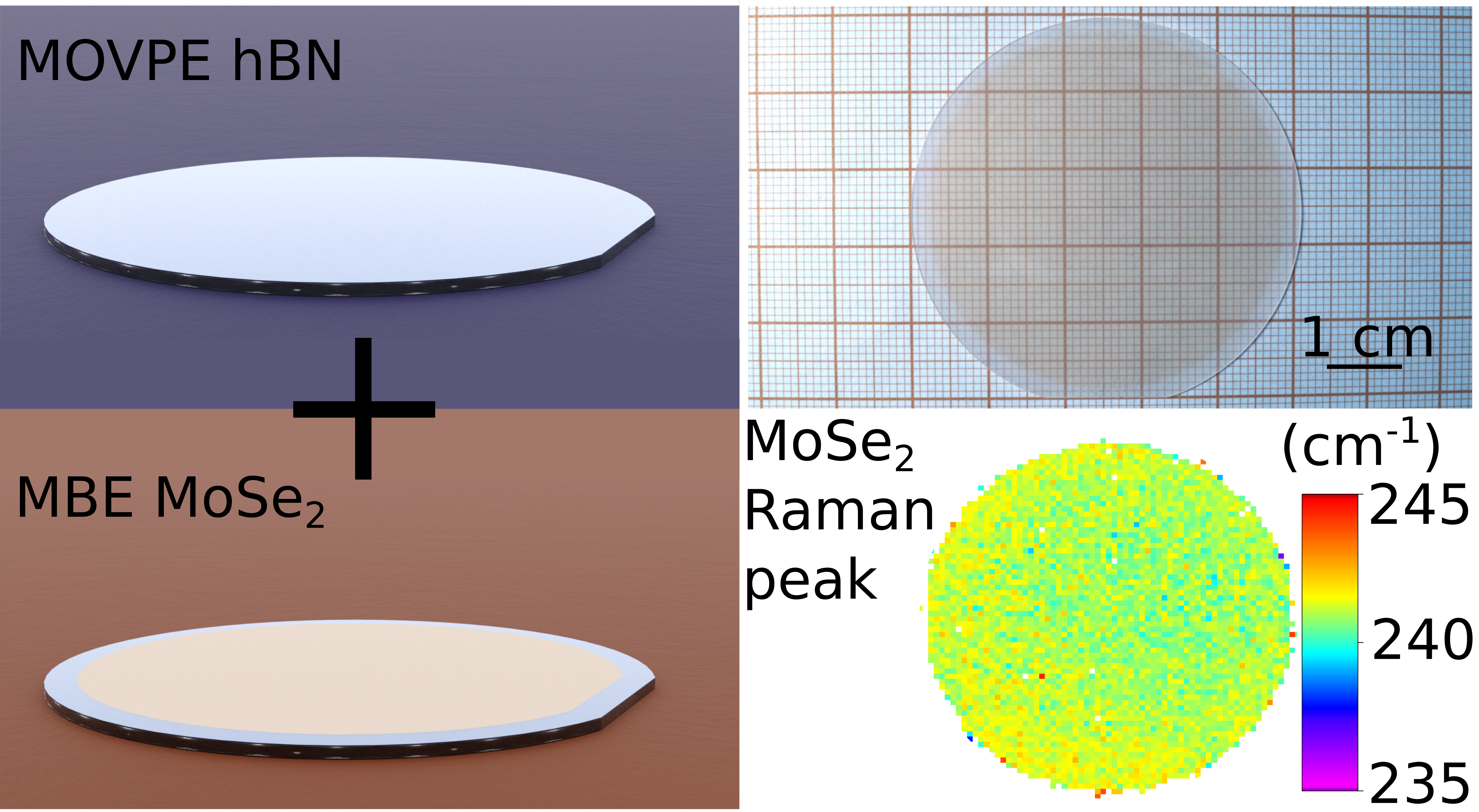}
\end{tocentry}


\begin{abstract}
Transition metal dichalcogenides (TMDs) are materials that can exhibit intriguing optical properties like a change of the bandgap from indirect to direct when being thinned down to a monolayer. Well-resolved narrow excitonic resonances can be observed for such monolayers, however only for materials of sufficient crystalline quality, so far mostly available in the form of micrometer-sized flakes. A further significant improvement of optical and electrical properties can be achieved by transferring the TMD on hexagonal boron nitride (hBN). To exploit the full potential of TMDs in future applications, epitaxial techniques have to be developed that not only allow to grow large-scale, high-quality TMD monolayers, but allow to perform the growth directly on large-scale epitaxial hBN. 
In this work we address this problem and demonstrate that MoSe$_2$ of high optical quality can be directly grown on epitaxial hBN on an entire two-inch wafer. We developed a combined growth theme for which hBN is first synthesized at high temperature by Metal Organic Vapor Phase Epitaxy (MOVPE)  and as a second step MoSe$_2$ is deposited on top by Molecular Beam Epitaxy (MBE) at much lower temperatures. We show that this structure exhibits excellent optical properties, manifested by narrow excitonic lines in the photoluminescence spectra. Moreover, the material is homogeneous on the area of the whole two-inch wafer, with only $\pm 0.14$ meV deviation of excitonic energy. Our mixed growth technique may guide the way for future large-scale production of high quality TMD/hBN heterostructures.
\end{abstract}
\section{Introduction}
Transition metal dichalcogenides (TMDs) - representatives of 2D layered materials, are intensively studied  as promising candidates for future realizations of optoelectronic devices \cite{Wang2012, Binder2017}, photodetectors \cite{Lopez-Sanchez2013, Mao2016, Wazir2020}, sensors \cite{Late2014, Zong2020, Awais2020}, energy and memory storages \cite{Bertolazzi2013, Yin2013, Yoshida2015} or transistors \cite{Radisavljevic2011, Yin2012, Zhang2012}. Such realizations are commonly demonstrated on micrometer-sized flakes obtained by mechanical exfoliation from bulk crystals and consecutively stacked to a heterostructure by time-consuming deterministic transfer processes \cite{Novoselov2005, Wang2012, Li2013, Li2014, Yi2015}. This approach allows to produce high quality samples in terms of electronic properties such as carrier mobility or conductivity \cite{Nasiri2019}.

\begin{figure}[h]
    \centering
    \includegraphics[width=1\textwidth]{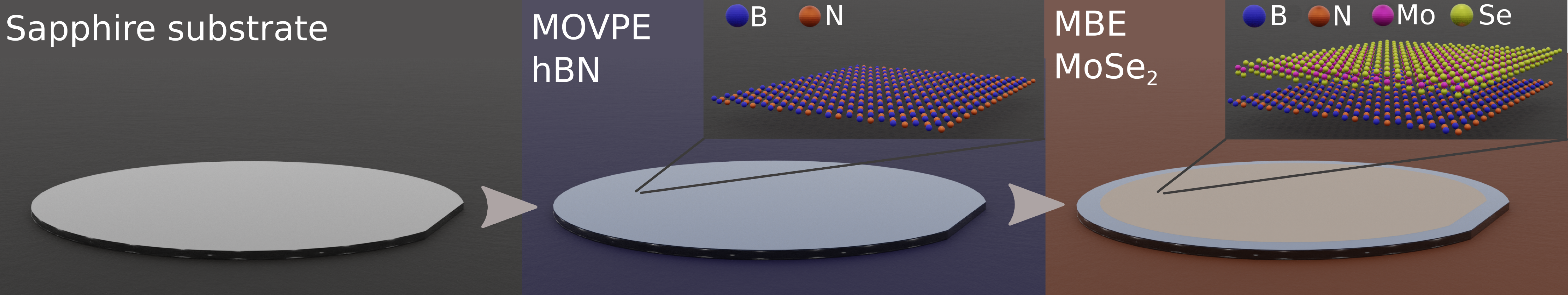}
    \caption{Schematic illustration of the combined growth approach. A two-inch sapphire wafer is used as a substrate for MOVPE hBN and MBE MoSe$_2$ growth.}
    \label{fig:sekwencja}
\end{figure}

Two-dimensional materials are, however, extremely sensitive to interlayer interactions and prone to environmental factors. Thus, it is important to deposit these materials on flat substrates with homogeneous dielectric properties and without dangling bonds. Hexagonal boron nitride (hBN) appears to be the perfect candidate for this purpose. The encapsulation in thin layers of hBN can significantly improve optical properties of the flakes and prolong their lifetime \cite{Cadiz, Courtade2018, Grzeszczyk2019, Han2019, Hayashida2020}.\\
The next necessary step towards the practical utilization of the properties of TMDs requires however, large-area sample synthesis. Significant efforts have been made to demonstrate wafer-scale growth \cite{Wang2020, Mandyam2020}. Approaches based on CVD (chemical vapour deposition) and PVD (physical vapour deposition) techniques result in promising large area monolayers \cite{Zhan2012, Duan2014, Liu2015, Jeon2015, Yu2017, Wang2020a}, but  excellent optical quality  could be only achieved after additional post-growth processing - by a sequence of transfer and encapsulation \cite{Delhomme2019, Shree2020}.\\
Epitaxial method like metalorganic vapour phase epitaxy (MOVPE) was also used to address the TMD scalability problem \cite{Marx2017} and it has been shown that it is possible to produce wafer-scale materials, however so far not with the same optical properties. Molecular beam epitaxy (MBE) appears to be a promising candidate for the development of large-area samples, as it allows to produce high purity materials with atomic precision \cite{Xenogiannopoulou2015, Ehlen2019, Wei2020, Wei2020a, Ohtake2020, He2020, Ogorzalek2020}. However, the optical quality of the samples is still far behind the best results obtained for mechanically exfoliated flakes. A way to overcome this obstacle is to combine TMD layers with hBN. Successful realizations of this approach such as MoSe$_2$ growth by MBE or CVD on initially exfoliated hBN flakes \cite{Yan2015, Pacuski2020} or an entirely grown hBN/MoS$_2$ heterostructure by CVD \cite{Fu2017} have been already presented.\\
Here, we present a method that combines two epitaxial techniques: MBE and MOVPE to provide large-scale crystals of high optical quality. We first use MOVPE to grow hBN layers of a few nanometres thickness on a two-inch sapphire wafer \cite{Dabrowska2020}. Subsequently, we use this sapphire/hBN wafer as a substrate to grow a monolayer of MoSe$_2$ by MBE. A schematic illustration of the growth sequence is presented in Figure \ref{fig:sekwencja}.

The quality of the TMDs is revealed by performing optical measurements such as photoluminescence and Raman spectroscopy. In particular, photoluminescence is a very sensitive probe of layer thickness, because materials like MoS$_2$, MoSe$_2$, WS$_2$, WSe$_2$ and MoTe$_2$ exhibit an indirect--direct band gap transition, which results in a bright and intense photoluminescence from monolayers, in contrast to photoluminescence from thicker samples \cite{Mak2010, Splendiani2010, Tonndorf2013, Zhang}.

Low-temperature photoluminescence spectra are a suitable indicator of the optical quality of MoSe$_2$, as excitonic lines can be resolved into two components, corresponding to the neutral A exciton and trion, only for high-quality samples \cite{Cadiz2016}. In our work, we show that the grown material is of excellent, homogeneous optical quality, confirmed by Raman spectroscopy and photoluminescence mapping of the two-inch wafer. We demonstrate the unique result of well-resolved excitonic lines at low temperatures on the whole wafer. 
The proposed approach provides a reliable template for the growth of high optical performance TMD layers by utilizing large area epitaxial hBN as a substrate.
Our method of combined MBE--MOVPE growth may pave the way for future applications of van der Waals heterostructures on the wafer scale.

\section{Results and discussion}
\subsection{Growth and characterization of hBN/MoSe$_2$ heterostructures}

Hexagonal boron nitride (hBN) layers were grown on 2-inch sapphire wafers by MOVPE.
The properties of the obtained hBN layers depend on a set of growth parameters including growth time, temperature, III-V ratio or the employed growth mode \cite{Pakua2019}. One of the key parameters in designing van der Waals heterostructures is the thickness of the hBN spacers. In this work we study the impact of hBN layer thickness on the properties of the subsequently grown MoSe$_2$ monolayer. To be able to obtain hBN layers of different thicknesses we employed two different growth modes: Continuous Flow Growth (CFG) and a two stage epitaxy to obtain three different hBN samples.
CFG growth mode yields ultrathin high quality hBN layers, but shows a self-limiting behaviour \cite{Nagashima1995}. Therefore, this mode cannot be used to obtain samples thicker than a few nanometres.
The grown material shows good optical and structural properties with thicknesses ranging from about $1$ nm to $6$ nm.  

To overcome the obstacle of limited thicknesses we employ the recently introduced two stage epitaxial growth \cite{Dabrowska2020}, which is a two-step mode for which first a high-quality CFG layer is grown followed by a pulsed growth step that requires an alternate switching of ammonia and TEB flows. With this mode it becomes possible to grow thicker hBN layers of exceptionally good structural quality (samples hBN1, hBN2) \cite{Dabrowska2020}. Samples grown using two stage epitaxy can reach thicknesses from several to tens of nanometers with a lattice constant close to the theoretical value of 3.33 Å \cite{Dabrowska2020}. The growth parameters of the studied samples are summarized in Table \ref{tab:growth_process}. 

\begin{table}[h]
\centering
\caption{MOVPE and MBE growth parametres of the samples presented in this study. In each MBE process about 1 ML MoSe$_2$ is depostied, but the growth rate is 3 times slower in case of the first process (A).}
\begin{tabular}{|c|c|c|}
\hline
\multicolumn{3}{|c|}{\textbf{MOVPE}}                           \\ \hline
\textbf{sample} & \textbf{thickness}   & \textbf{growth mode}  \\ \hline
hBN1            & 13.6 nm              & two stage epitaxy                   \\ \hline
hBN2            & 3.5 nm               & two stage epitaxy                   \\ \hline
hBN3            & 1.5 nm               & CFG                   \\ \hline
\multicolumn{3}{|c|}{\textbf{MBE}}                             \\ \hline
\textbf{sample} & \multicolumn{2}{c|}{\textbf{time of growth}} \\ \hline
MoSe$_2$A          & \multicolumn{2}{c|}{15 h}                    \\ \hline
MoSe$_2$B          & \multicolumn{2}{c|}{5 h}                     \\ \hline
MoSe$_2$C          & \multicolumn{2}{c|}{5x1 h}                   \\ \hline
\end{tabular}
\label{tab:growth_process}
\end{table}

\begin{figure}[h]
    \centering
    \includegraphics[width=1\textwidth]{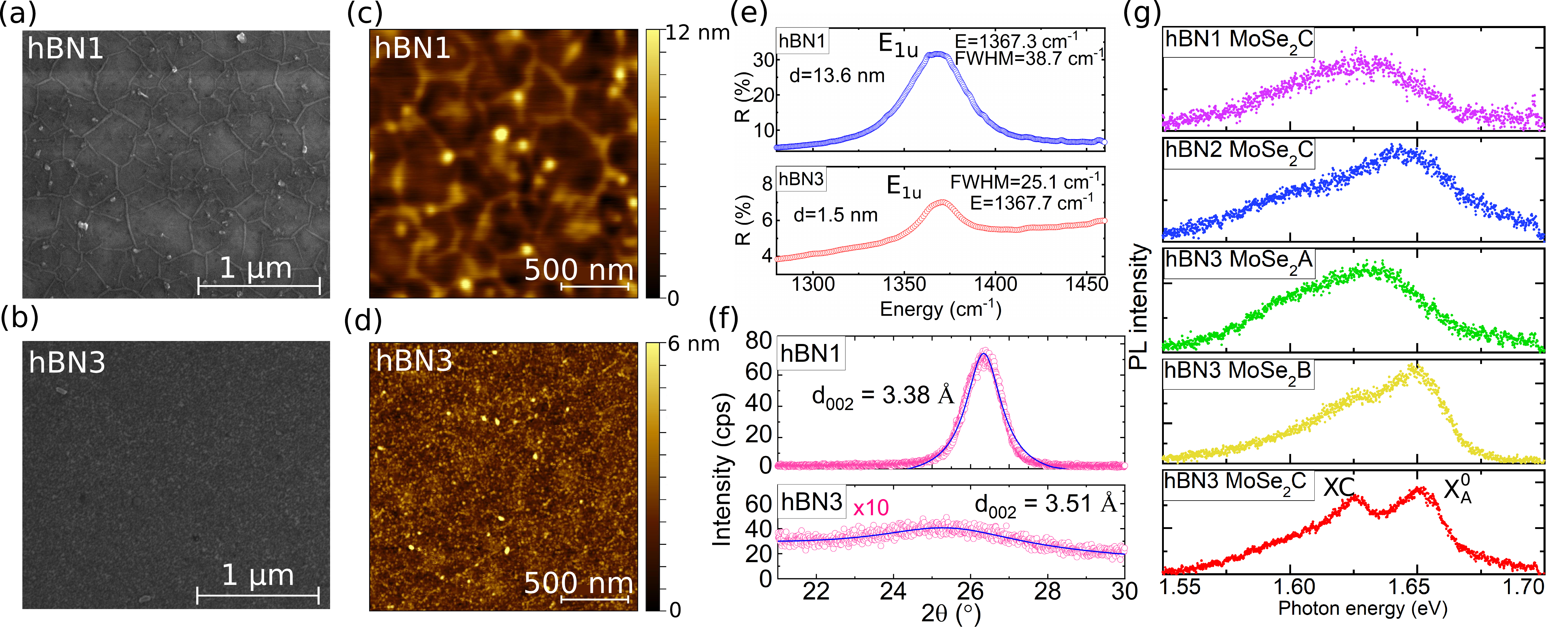}
    \caption{Characterization of hBN substrates obtained in different growth modes. SEM images of (a) hBN1, (b) hBN3; AFM images of (c) hBN1, (d) hBN3 show the topography of the samples -- the thinner the hBN layer, the smaller the precipitations and wrinkles. (e) FTIR spectra for hBN samples: E$_{1u}$ mode with corresponding energy and thickness. (f) Results of a $2\theta$/$\omega$ scan of the samples with extracted interplanar spacing. (g) Low-temperature photoluminescence spectra obtained for various hBN substrates and MBE growth processes.}
    \label{fig:zbiorczy}
\end{figure}

Figure \ref{fig:zbiorczy} presents SEM and AFM images showing the morphology of the thickest (hBN1) and thinnest (hBN3) samples. The MOVPE growth of hBN at high temperature is unavoidably connected to the appearance of characteristic wrinkles on the material. Their presence can be attributed to the post-growth cooling process and different thermal expansion coefficients of sapphire and hBN. This wrinkle pattern shows a larger mesh size in the case of thicker samples. Other typical objects on the surface of the material are three-dimensional precipitates. Their appearance is most probably the result of the formation of out-of-plane nucleation centres. There is a strong dependence of the precipitates size. The thinner the sample, the smaller the three-dimensional precipitates.  

To measure the layer thickness and to further characterize the hBN, we performed infrared reflectance spectra measurements using a FTIR microscope and TEM imaging (see supporting information Figure S1). By analysing the spectra within the Dynamic Dielectric Function approximation we were able to determine the thicknesses of the hBN layers (Figure \ref{fig:zbiorczy}(e)) \cite{Dabrowska2020}. The characteristic peak in the FTIR spectra corresponds to the E$_{1u}$ mode of hBN around $1367$ cm$^{-1}$ confirming the formation of a sp$^2$-BN layer \cite{Geick1966}.\\

X-ray diffraction (XRD) was used to investigate the structural quality of the samples. The XRD results are presented in Figure \ref{fig:zbiorczy}(f). The peak position of around $26.3^{\circ}$ for sample hBN1 corresponds to hBN with some turbostratic addition. The peak observed for the thinner sample is less intense and broader,  and its angular position of $24.7^{\circ}$ suggest mostly turbostratic stacking \cite{Kobayashi2008}.

The hBN samples with various thickness were used as substrates for the consecutive growth of MoSe$_2$ by MBE.
Various growth processes were performed to optimize monolayer formation. We studied the effect of different growth times (growth rates) as well as using one or several annealing steps.

Figure \ref{fig:zbiorczy}(g) depicts low temperature photoluminescence spectra for MoSe$_2$ layers grown with various MBE processes (MoSe$_2$A, MoSe$_2$B, MoSe$_2$C) and on various epitaxial hBN substrates (hBN1, hBN2, hBN3). At low temperatures, the photoluminescence spectrum of MoSe$_2$ unveils additional information about the optical quality of the material. For high-quality samples, the excitonic line separates into two well-resolved peaks corresponding to the A exciton and charged exciton (trion). 

Such spectra were mostly observed for mechanically exfoliated material onto selected substrates, \cite{Cadiz,Ross2013,Zhang2018,Ye2018} but recently high optical quality samples were also obtained for MBE grown MoSe$_2$ on exfoliated hBN \cite{Pacuski2020}.

The analysis of low-temperature PL spectra of different samples (Figure \ref{fig:zbiorczy}(g)) allows us to optimize  the process of heteroepitaxial growth. The growth parameters for the MoSe$_2$A sample were adopted from previous experiments on exfoliated hBN \cite{Pacuski2020}. This  starting point of parameters resulted in a broad spectrum, which can indicate the formation of more than a single atomic layer.
The reduction of the growth time for sample MoSe$_2$B resulted in a notable improvement of the optical quality of the material and a visible separation of excitonic lines into two peaks (A exciton and trion). A further introduction of several annealing steps into the process resulted in an even clearer separation of the exciton and trion peaks for sample MoSe$_2$C. This growth process was performed on three different hBN substrates (hBN1, hBN2, hBN3) during the same growth run. It can be seen that the thinner the hBN layer, the higher the optical quality of the material.

\begin{figure}[h]
    \centering
    \includegraphics[width=0.7\textwidth]{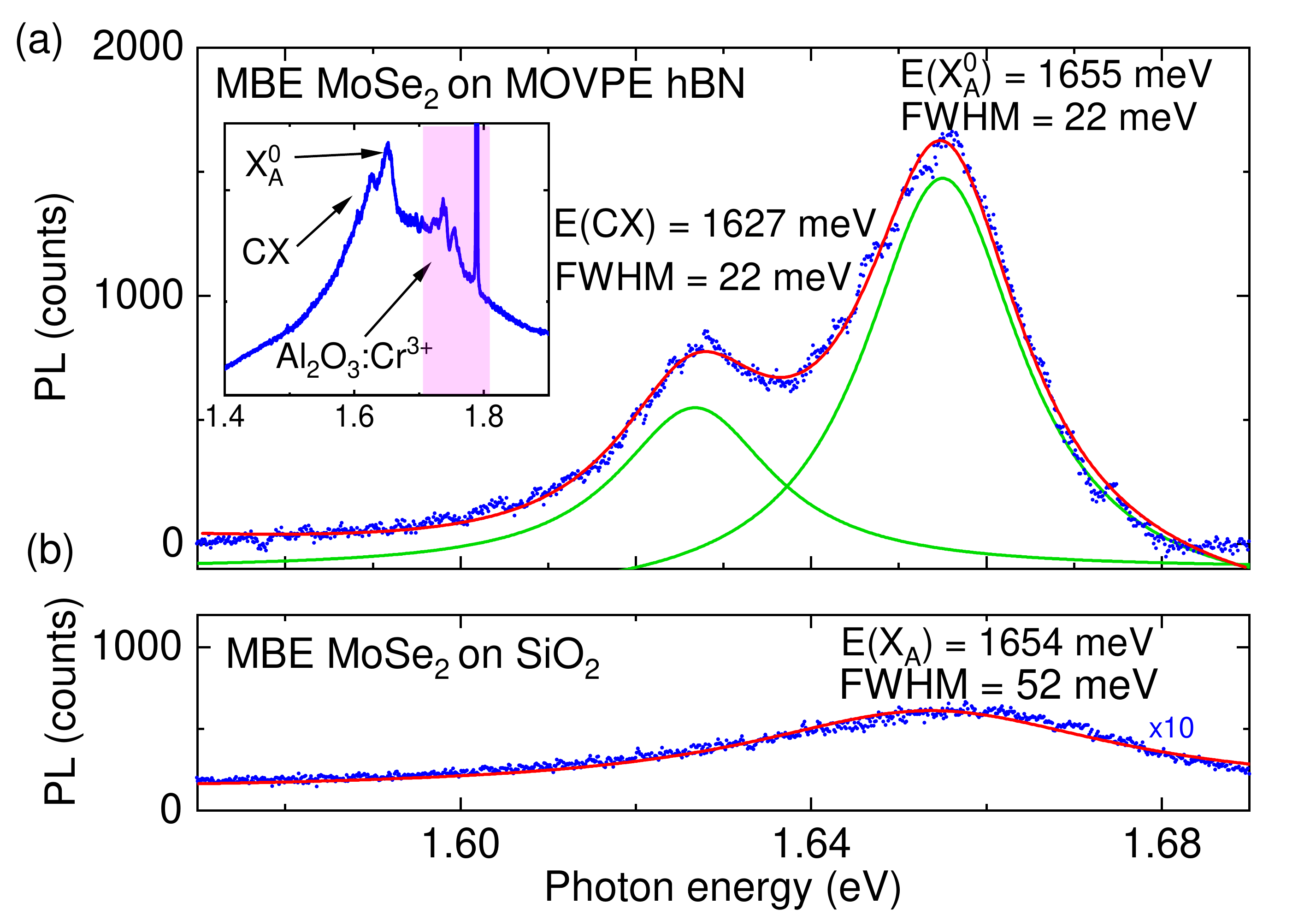}
    \caption{Low temperature (5 K) photoluminescence spectra of MoSe$_2$ grown by MBE on a MOVPE grown hBN compared to material grown by the same MBE process directly on a silicone dioxide substrate. (a) Photoluminescence spectrum of MoSe$_2$ grown using optimized MBE and MOVPE processes. The lines of the neutral A exciton (X$_A^0$) and charged exciton (XC) can be resolved. Blue dots show the data, while red and green curves show results of Lorentzian fits. Inset -- photoluminescence spectrum measured  in a larger spectral range showing an additional background which emerges from the defects in hBN. (b) photoluminescence spectrum obtained for MBE MoSe$_2$ grown on a SiO$_2$ substrate. Excitonic lines are not resolved and the line width of the neutral exciton around $1654$ meV is significantly larger (FWHM $52$ meV) as compared to the material grown on MOVPE hBN.}
    \label{fig:widma_PL_5K}
\end{figure}

 The inset in Figure \ref{fig:widma_PL_5K}(a) shows a typical spectrum obtained for MoSe$_2$C processes grown on hBN3 MOVPE samples in a larger spectral range. The additional background stems from the photoluminescence of defects in the hBN layer. The peaks around $1.7-1.8$ eV correspond to the presence of chromium in the sapphire substrates \cite{Macfarlane1963}. 
 Figure \ref{fig:widma_PL_5K}(a) depicts two distinct peaks typical for MoSe$_2$ corresponding to the neutral A exciton and trion with the subtracted background. Peaks are visible at energies of $1655$ meV and $1627$ meV, respectively. These results correspond well with the excitonic complexes reported in previous works \cite{Ross2013,Wang2015,Cadiz,Wierzbowski2017}. The peak width for the A exciton and trion is $22$ meV, which is significantly smaller than the peak width measured for the MoSe$_2$ grown on a SiO$_2$ substrate \ref{fig:widma_PL_5K}(b) and only a factor of two larger than for MoSe$_2$ grown on flakes exfoliated from high-quality bulk hBN \cite{Pacuski2020}.   

\begin{figure}[h]
\includegraphics[width=0.75\textwidth]{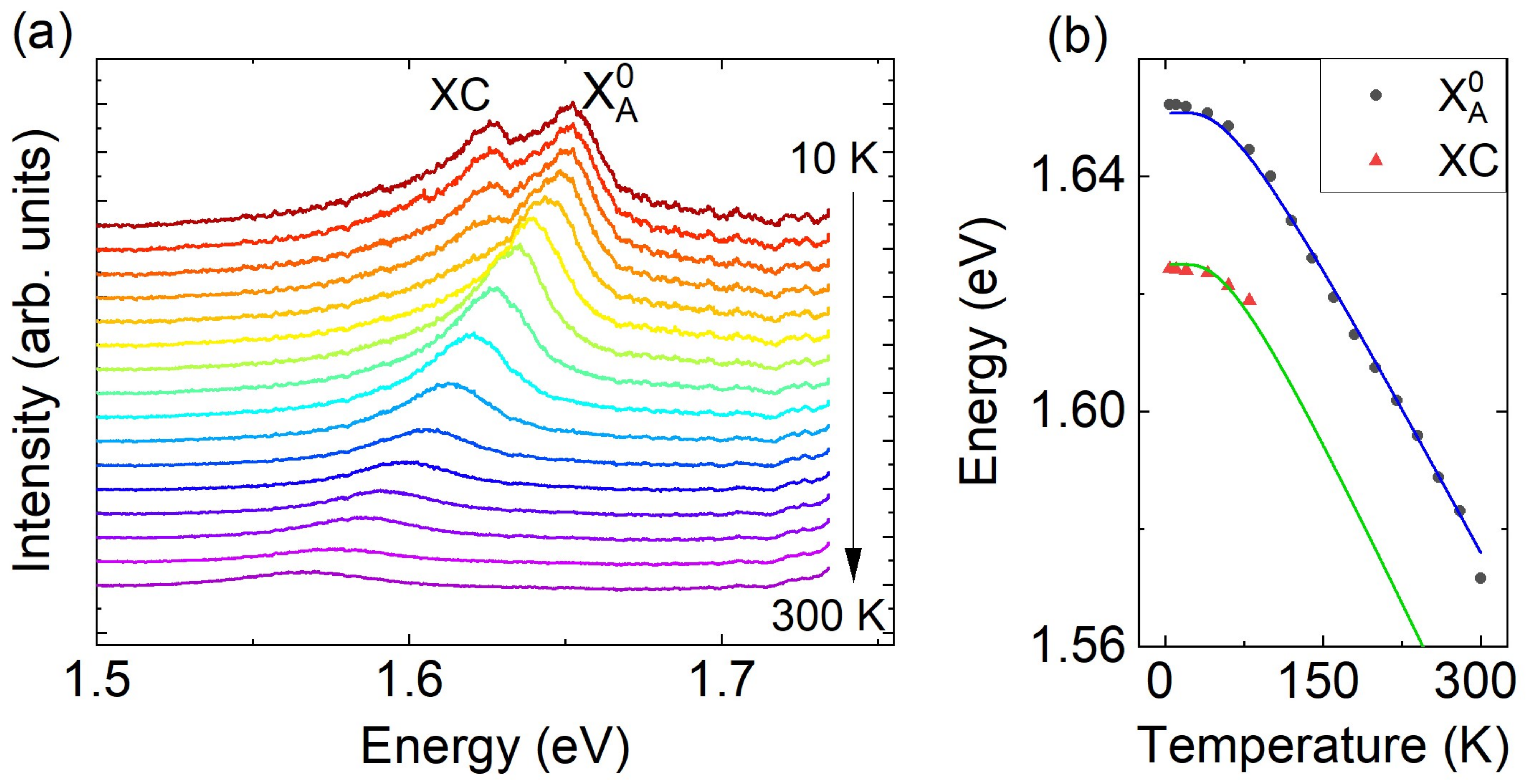}
\centering
\caption{(a) Temperature dependence of the photoluminescence of a MoSe$_2$/hBN sample. The spectra were shifted vertically for clarity. At T=$10$ K the spectra consists of both, the charged and neutral excitonic lines. The charged exciton contribution disappears at around $100$ K. (b) Temperature dependence of the energy of the charged and neutral exciton. The blue and green curves present a typical temperature dependence of the excitonic lines of MoSe$_2$. The parameters of the curves were taken from \cite{Ross2013}.}
\label{fig:zaleznosc_temp}
\end{figure}

The temperature dependence of the excitonic resonances observed in our samples is typical for monolayer MoSe$_2$. Figure \ref{fig:zaleznosc_temp}(a) depicts the evolution of the photoluminescence of the neutral exciton A and trion as a function of temperature (laser excitation wavelength $\lambda = 532$). At low temperatures we observe a binding energy of about $30$ meV. The trion signal abruptly decays at around $100$ K. Both peaks shift towards lower energies as the temperature rises, due to the reduction of the band gap, shown in Figure \ref{fig:zaleznosc_temp}(b)\cite{Ross2013, Godde2016}. The blue and green lines are not fits to our data, but show the typical band-gap energy dependence with parameters taken from \cite{Ross2013} indicating that our material indeed behaves like typical MoSe$_2$ without significant influence of stress due to the interaction with hBN layer.

\subsection{Wafer-scale growth of a hBN/MoSe$_2$ heterostrucutre}
\begin{figure}[]
    \centering
    \includegraphics[width=1\textwidth]{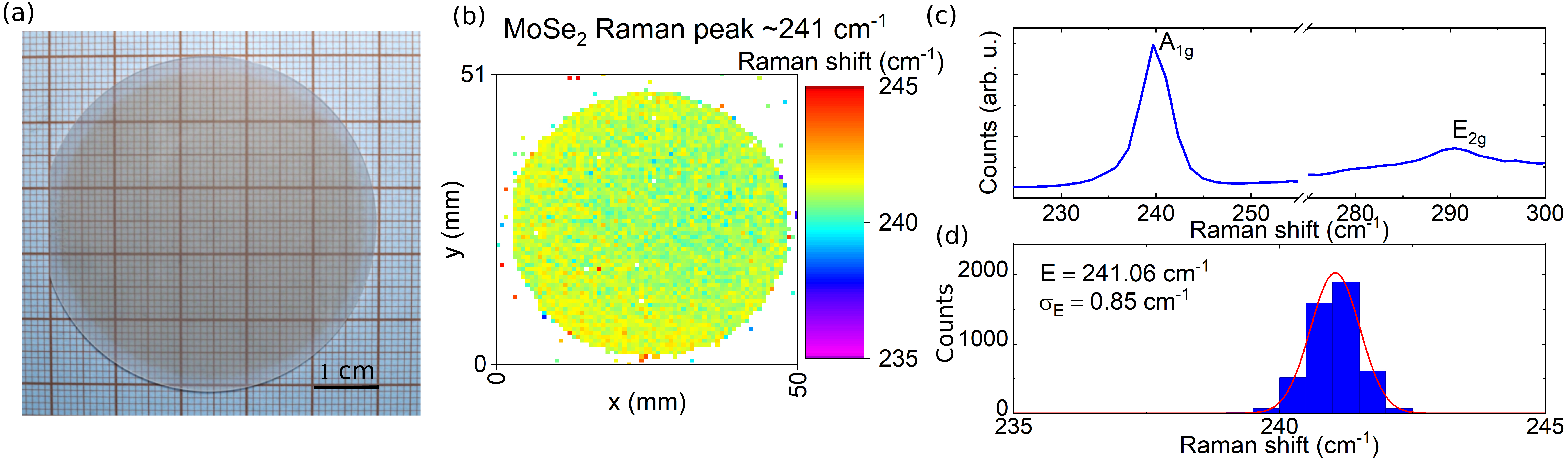}
    \caption{(a) Picture of an epitaxially grown two-inch hBN/MoSe$_2$ van der Waals heterostructure. The visibly darker reddish color indicates MoSe$_2$ coverage. (b) Peak energy of the A$_{1g}$ mode obrained by Raman spectroscopy mapping at room temperature. (c) Typical Raman spectrum of the two characteristic MoSe$_2$ peaks. (d) Distribution of the peak energy with a mean value of $E_{A_{1g}} = 241.06$ cm$^{-1}$ and standard deviation $\sigma_{E}=0.85$ cm$^{-1}$. The results from the Raman map analysis show that the whole area of the sample is uniformly covered by a single layer of the MoSe$_2$.}
    \label{fig:zdjecie_raman}
\end{figure}

To further prove that our method can be scaled up, we chose the best MOVPE and MBE processes and grew the layers on an entire two-inch sapphire substrate. Figure \ref{fig:zdjecie_raman}(a) presents a picture of the 2-inch wafer after hBN and MoSe$_2$ growth. The material can be observed with the naked eye. The MoSe$_2$ layer covers the sample in a circular, visibly darker, brownish area. The transparent, uncovered outer ring is the result of applying a frame that holds the  substrate in the MBE chamber. It also allows to examine the properties of hBN after the MBE growth.

The sample was further studied using Raman spectroscopy.
Figure \ref{fig:zdjecie_raman}(b) presents the outcome of room temperature Raman spectroscopy mapping of the whole wafer. The out-of-plane A$_{1g}$ mode of MoSe$_2$ at $241$ cm$^{-1}$ was chosen for the map analysis, as it is more intense than the in-plane E$_{2g}$ peak (Figure \ref{fig:zdjecie_raman} (c)). The presented map confirms that the material indeed uniformly covers the whole area of the sample. Raman spectroscopy is also an excellent tool to assess the number of MoSe$_2$ layers \cite{Tonndorf2013, Zhang2019}. As compared to the bulk material the out-of-plane mode softens (redshifts) while the in-plane mode stiffens (blueshifts) with decreasing thickness. A typical Raman spectrum for this sample is presented in Figure \ref{fig:zdjecie_raman}(c). The highlighted Raman peak positions indicate the presence of a single layer of MoSe$_2$. While the peak energy can also vary due to different substrates and exfoliation/growth techniques \cite{Matte2011, MuiPoh2018}, the mean value of the A$_{1g}$ peak energy of $E_{A_{1g}}=241.06$ cm$^{-1}$ for the whole area of the sample with a standard deviation as small as $\sigma_{E}=0.85$ cm$^{-1}$ indeed indicates the formation of a monolayer material. The distribution of the peak energy quantitatively depicted in Figure \ref{fig:zdjecie_raman}(d) additionally shows the homogeneity of the material.    

\begin{figure}[h]
    \includegraphics[width=0.9\textwidth]{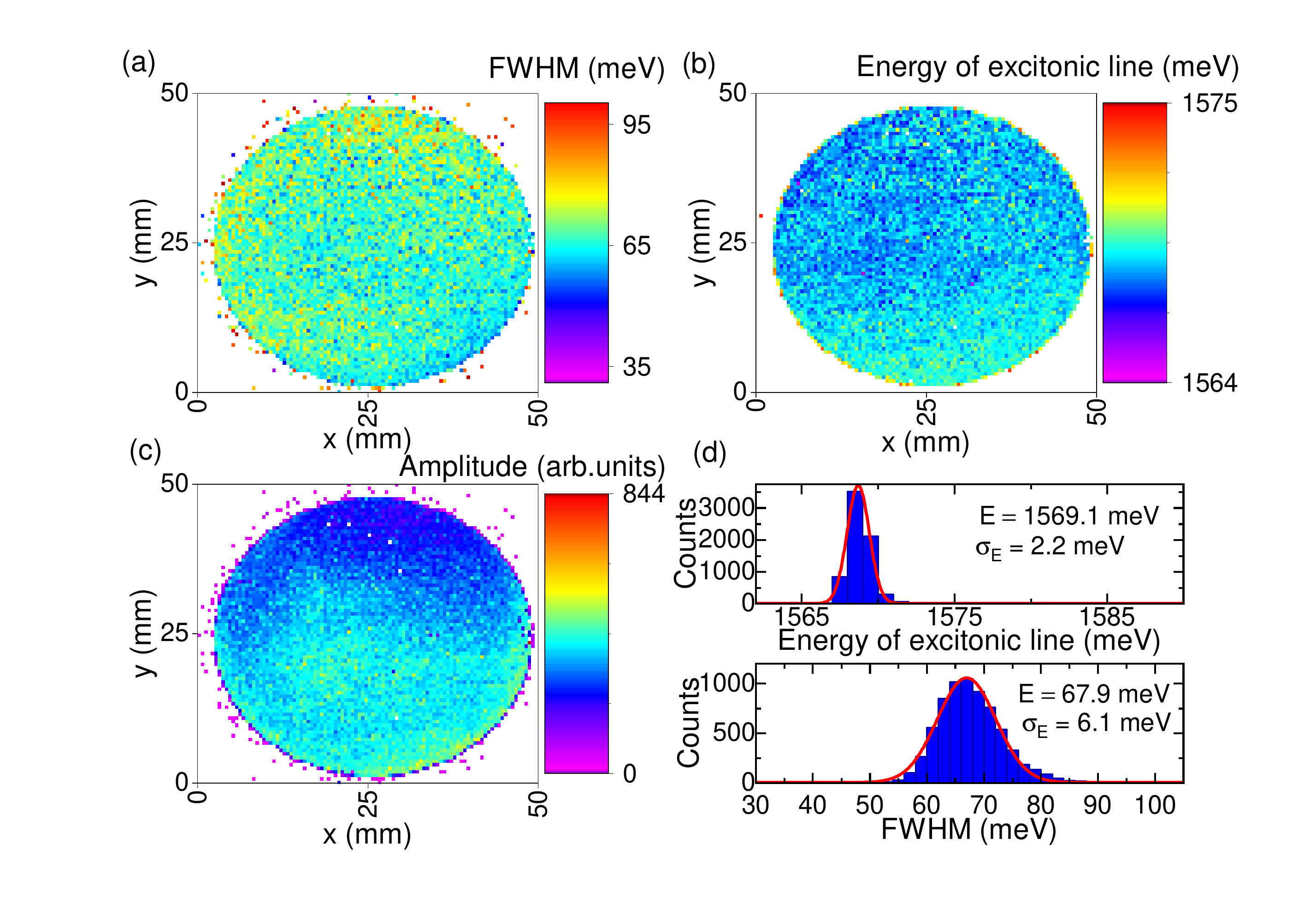}
    \centering
    \caption{Photoluminescence mapping at room temperature ($500$ $\mu$m step, $633$ nm laser). Panels (a),(b),(c) show results of Lorentzian fits to the obtained spectra. A peak width of $67.9\pm6.1$ meV and an energy of $1569.1\pm 2.1$ meV for the excitonic line were obtained. The low standard deviation indicates sample homogeneity across the whole two-inch wafer. (d) Histograms of the excitonic line energy (top) and width (bottom).}
    \label{fig:mapa_PL}
\end{figure}

Further evidence for the presence of monolayer MoSe$_2$ can be provided by photoluminescence mapping. The bandgap becomes direct in the monolayer limit as there are no more vdW interlayer interactions \cite{Yun2012} rendering photoluminescence a very sensitive tool with regard to thickness fluctuations. At room temperature, such a spectrum consists of one distinct peak corresponding to the neutral A exciton with an energy of $1.57$ eV for monolayer and $1.54$ eV for bilayer MoSe$_2$ \cite{Tonndorf2013}.
Figure \ref{fig:mapa_PL}(a)-(d) depicts the results of PL mapping of the whole wafer at room temperature. 
The extracted peak width (Figure \ref{fig:mapa_PL}~(a)) is uniform across the sample with a mean value of $68\pm 6$ meV, which is very low for spectra obtained at room temperature \cite{Ross2013}. The energy of the excitonic line does not vary much between different positions on the sample. Its mean value of $1569.1 \pm 2.2$ meV corresponds perfectly to previously reported results for monolayer MoSe$_2$ \cite{Tonndorf2013}. Another important factor in determining the thickness of the material is the photoluminescence intensity. It decreases usually by one order of magnitude with every additional layer. In our case, the intensity is homogeneous over the whole area of the sample (Figure~\ref{fig:mapa_PL}~(c)). Transmission Electron Microscope (TEM) characterization of the final sample is in agreement with the optical studies indicating the formation of a MoSe$_2$ monolayer (see supporting information Figure S1). Further characterization of the final sample morphology is provided by AFM (see supporting information Figure S2).

\begin{figure}[h]
\includegraphics[width=0.89\textwidth]{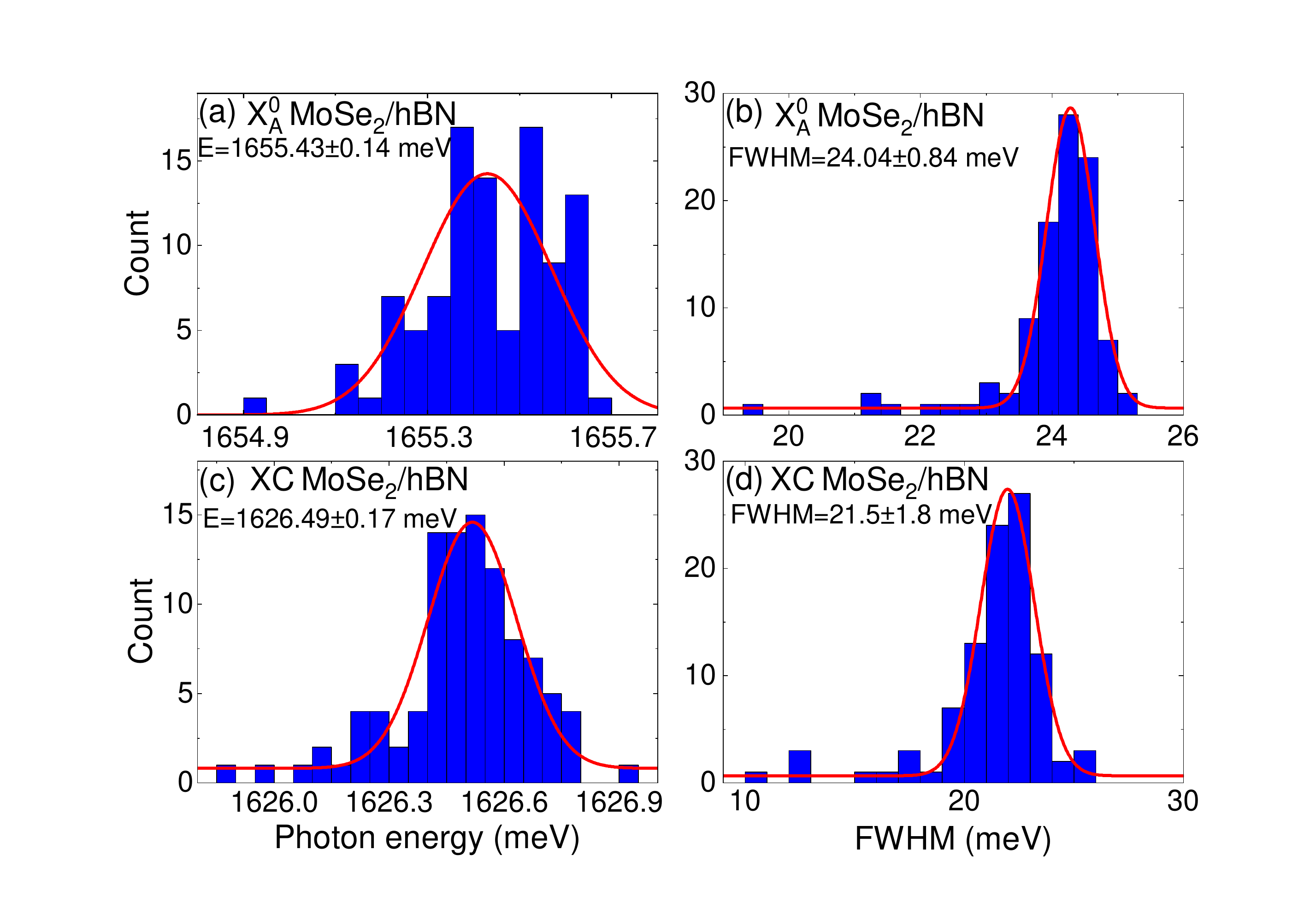}
\centering
 \caption{Photoluminescence peaks statistics at T= $5$~K. The spectra were measured along a line of 1~cm. The histograms show the result of Lorentzian fits to the neutral (X$^0_A$) and charged exciton (XC). (a) distribution of the energy of X$^0_A$. (b) Distribution of peak width of X$^0_A$. (c) Distribution of the energy of XC. (d) Distribution of peak width of XC.}
    \label{fig:histogramy}
\end{figure}

Figure \ref{fig:histogramy}(a)-(d) depicts the distributions of the peak energy of the A exciton and trion as well as their widths acquired from mapping the sample at liquid helium temperature. Due to the limited size of the cryostat window, we measured $100$ spectra uniformly along a straight line ($1$ cm). A mean energy of $1655.43 \pm 0.14$ meV and $1626.49 \pm 0.17$ meV was extracted for the peaks corresponding to the A exciton and trion, which confirms the large homogeneity of the sample. This aspect is especially important for any future industrial application of TMDs requiring reproducible results, such as a well-defined band gap energy. 
 
\newpage
\section{Conclusions}
We present a method to produce wafer-scale monolayer TMD samples of high optical quality. Our method combines two epitaxial growth methods: MOVPE, which enables us to grow hBN on a sapphire substrate and MBE which was used to cover the so-grown hBN layer with a monolayer of MoSe$_2$. A thorough analysis of the optical properties examined by photoluminescence and Raman spectroscpy mapping reveals a large homogeneity of the epitaxially grown van der Waals heterostructure on the whole two-inch wafer. We also demonstrate that well-resolved and narrow excitonic peaks at low temperature can be achieved even for such large heteroepitaxial samples. The uniform energy distributions with small standard deviations indicate an excellent homogeneity and repeatability, which is very difficult to achieve with other techniques like mechanical exfoliation. Our results constitute a large step towards a possible industrial implementation of van der Waals heterostructures by enabling wafer-scale production with precisely defined properties. 

\section{Experimental section}
\subsection{hBN growth by MOVPE} 
An Aixtron Close Coupled Showerhead 3×2” system with ARGUS thermal mapping device was used to grow hBN samples. 2'' sapphire wafers purchased from ProCrystal were used as a substrate. Triethylboron (TEB) and ammonia (NH$_3$) were used as boron and nitrogen precursors, with H$_2$ as carrier gas.
Two growth modes were applied to produce the samples discussed in this work:\\
- Continuous Flow Growth (sample hBN3): precursors are simultaneously injected to the reactor at the same time at high temperatures around $1300^\circ$C with supersaturation of ammonia \cite{Pakua2019,Dabrowska2020}.Such a mode exhibit a self-limiting behavior. The obtained samples show excellent optical quality, but cannot be grown thicker than $\sim$6 nm.\\
- Two Step Epitaxy (samples hBN1, hBN2) \cite{Dabrowska2020}: The growth of a CFG buffer is followed by a pulsed growth step requiring the switching of ammonia and TEB flows. The obtained samples exhibit excellent structural quality and this way thicker layers than $\sim$6 nm can be grown.
\subsection{MoSe$_2$ growth by MBE} 
A MBE machine fabricated by SVT Associates was used to grow MoSe$_2$ layers. To optimize the process, three hBN samples (hBN1, hBN2, hBN3) were used as substrates for subsequent MBE growth.\\
- Sample MoSe$_2$A was grown in 3 growth steps -- each of which lasted for 5 hours and was followed by $20$ min of heating up to $700^{\circ}$C and $10$ min annealing in Se flux.\\
- Sample MoSe$_2$B was grown just for 5 hours (with 3 times higher growth rate) with only one annealing step at the very end of the process. This time the temperature of the sample was slowly increased up to $700^{\circ}$C and the annealing in Se flux lasted for two hours. \\
- Sample MoSe$_2$C was grown in 5 consecutive cycles. Each step contained one hour of growth at $300^{\circ}$C, a slow heating up to $760^{\circ}$C and $10$ min annealing in Se flux. The last annealing was however significantly longer and lasted for 2 hours.\\
The selection of growth parameters resulted from previous experiences with MoSe$_2$ growth on exfoliated hBN as well as optimization to achieve the best optical quality of the samples.
\subsection{Characterization of the samples}
For low temperature ($5$ K) measurements, samples were placed in liquid helium continuous flow cryostat (MicrostatHires Oxford Instruments). Luminescence spectra at $\sim$5 K were measured with a HORIBA T64000 spectrometer with Nd:YAG $532$ nm continuous wave laser.\\
Raman and photoluminescence mapping of the sample at room temperature was performed with Renishaw inVia spectrometers. The material was excited by a $\lambda$=$532$ nm laser with a power of $320$ $\mu$W and a 20x objective (NA 0.4) for Raman mapping and by $\lambda$=$633$ nm laser, $550$ $\mu$W laser power and a 20x objective (NA 0.4) for photoluminescence mapping.\\
XRD measurements were performed using a Panalytical X’pert diffractometer equipped with Cu X-ray tube.\\
FTIR reflectance spectra were obtained by Thermo Fischer Scientific iS50, Nicolet Continuum and analyzed within the Dynamic Dielectric Functin approximation.\\
SEM images were acquired using FEI Helios NanoLab 600 and AFM images were acquired by Digital Instruments MMAFM-2 atomic force microscopy. \\
The morphology and structure of MoSe$_2$/BN on sapphire were investigated using a FEI Titan 80-300 transmission electron microscope operating at 300 kV, equipped with an image corrector. The electron transparent cross-sections of the heterostructure were cut using a focused gallium ion beam with a Helios Nanolab 600 focused ion beam (FIB) system. Prior to the ion cutting, a gold cap was deposited by magnetron sputtering, which was necessary to eliminate sample charging during TEM lamella preparation.

\begin{acknowledgement}

This work was supported by the National Science Centre, Poland under decisions 2019/33/B/ST5/02766, 2017/27/B/ST5/02284 and 2020/39/D/ST7/02811.

\end{acknowledgement}

\bibliography{biblio3}

\providecommand{\latin}[1]{#1}
\makeatletter
\providecommand{\doi}
  {\begingroup\let\do\@makeother\dospecials
  \catcode`\{=1 \catcode`\}=2 \doi@aux}
\providecommand{\doi@aux}[1]{\endgroup\texttt{#1}}
\makeatother
\providecommand*\mcitethebibliography{\thebibliography}
\csname @ifundefined\endcsname{endmcitethebibliography}
  {\let\endmcitethebibliography\endthebibliography}{}
\begin{mcitethebibliography}{67}
\providecommand*\natexlab[1]{#1}
\providecommand*\mciteSetBstSublistMode[1]{}
\providecommand*\mciteSetBstMaxWidthForm[2]{}
\providecommand*\mciteBstWouldAddEndPuncttrue
  {\def\EndOfBibitem{\unskip.}}
\providecommand*\mciteBstWouldAddEndPunctfalse
  {\let\EndOfBibitem\relax}
\providecommand*\mciteSetBstMidEndSepPunct[3]{}
\providecommand*\mciteSetBstSublistLabelBeginEnd[3]{}
\providecommand*\EndOfBibitem{}
\mciteSetBstSublistMode{f}
\mciteSetBstMaxWidthForm{subitem}{(\alph{mcitesubitemcount})}
\mciteSetBstSublistLabelBeginEnd
  {\mcitemaxwidthsubitemform\space}
  {\relax}
  {\relax}

\bibitem[Wang \latin{et~al.}(2012)Wang, Kalantar-Zadeh, Kis, Coleman, and
  Strano]{Wang2012}
Wang,~Q.~H.; Kalantar-Zadeh,~K.; Kis,~A.; Coleman,~J.~N.; Strano,~M.~S.
  {Electronics and optoelectronics of two-dimensional transition metal
  dichalcogenides}. \emph{Nature Nanotechnology} \textbf{2012}, \emph{7},
  699--712\relax
\mciteBstWouldAddEndPuncttrue
\mciteSetBstMidEndSepPunct{\mcitedefaultmidpunct}
{\mcitedefaultendpunct}{\mcitedefaultseppunct}\relax
\EndOfBibitem
\bibitem[Binder \latin{et~al.}(2017)Binder, Withers, Molas, Faugeras,
  Nogajewski, Watanabe, Taniguchi, Kozikov, Geim, Novoselov, and
  Potemski]{Binder2017}
Binder,~J.; Withers,~F.; Molas,~M.~R.; Faugeras,~C.; Nogajewski,~K.;
  Watanabe,~K.; Taniguchi,~T.; Kozikov,~A.; Geim,~A.~K.; Novoselov,~K.~S.;
  Potemski,~M. {Sub-bandgap Voltage Electroluminescence and
  Magneto-oscillations in a WSe$_2$ Light-Emitting van der Waals
  Heterostructure}. \emph{Nano Letters} \textbf{2017}, \emph{17},
  1425--1430\relax
\mciteBstWouldAddEndPuncttrue
\mciteSetBstMidEndSepPunct{\mcitedefaultmidpunct}
{\mcitedefaultendpunct}{\mcitedefaultseppunct}\relax
\EndOfBibitem
\bibitem[Lopez-Sanchez \latin{et~al.}(2013)Lopez-Sanchez, Lembke, Kayci,
  Radenovic, and Kis]{Lopez-Sanchez2013}
Lopez-Sanchez,~O.; Lembke,~D.; Kayci,~M.; Radenovic,~A.; Kis,~A.
  {Ultrasensitive photodetectors based on monolayer MoS$_2$}. \emph{Nature
  Nanotechnology} \textbf{2013}, \emph{8}, 497--501\relax
\mciteBstWouldAddEndPuncttrue
\mciteSetBstMidEndSepPunct{\mcitedefaultmidpunct}
{\mcitedefaultendpunct}{\mcitedefaultseppunct}\relax
\EndOfBibitem
\bibitem[Mao \latin{et~al.}(2016)Mao, Yu, Wang, Zhang, Wang, Shao, and
  Jie]{Mao2016}
Mao,~J.; Yu,~Y.; Wang,~L.; Zhang,~X.; Wang,~Y.; Shao,~Z.; Jie,~J. {Ultrafast,
  Broadband Photodetector Based on MoSe$_2$/Silicon Heterojunction with
  Vertically Standing Layered Structure Using Graphene as Transparent
  Electrode}. \emph{Advanced Science} \textbf{2016}, \emph{3}\relax
\mciteBstWouldAddEndPuncttrue
\mciteSetBstMidEndSepPunct{\mcitedefaultmidpunct}
{\mcitedefaultendpunct}{\mcitedefaultseppunct}\relax
\EndOfBibitem
\bibitem[Wazir \latin{et~al.}(2020)Wazir, Liu, Ding, Wang, Ye, Lingling, Lu,
  Wei, and Zou]{Wazir2020}
Wazir,~N.; Liu,~R.; Ding,~C.; Wang,~X.; Ye,~X.; Lingling,~X.; Lu,~T.; Wei,~L.;
  Zou,~B. {Vertically Stacked MoSe$_2$/MoO$_2$ Nanolayered Photodetectors with
  Tunable Photoresponses}. \emph{ACS Applied Nano Materials} \textbf{2020},
  \emph{3}, 7543--7553\relax
\mciteBstWouldAddEndPuncttrue
\mciteSetBstMidEndSepPunct{\mcitedefaultmidpunct}
{\mcitedefaultendpunct}{\mcitedefaultseppunct}\relax
\EndOfBibitem
\bibitem[Late \latin{et~al.}(2014)Late, Doneux, and Bougouma]{Late2014}
Late,~D.~J.; Doneux,~T.; Bougouma,~M. {Single-layer MoSe$_2$ based NH3 gas
  sensor}. \emph{Applied Physics Letters} \textbf{2014}, \emph{105},
  233103\relax
\mciteBstWouldAddEndPuncttrue
\mciteSetBstMidEndSepPunct{\mcitedefaultmidpunct}
{\mcitedefaultendpunct}{\mcitedefaultseppunct}\relax
\EndOfBibitem
\bibitem[Zong \latin{et~al.}(2020)Zong, Li, Chen, Liu, Li, Ruan, and
  Mao]{Zong2020}
Zong,~B.; Li,~Q.; Chen,~X.; Liu,~C.; Li,~L.; Ruan,~J.; Mao,~S. {Highly Enhanced
  Gas Sensing Performance Using a 1T/2H Heterophase MoS$_2$ Field-Effect
  Transistor at Room Temperature}. \emph{ACS Applied Materials and Interfaces}
  \textbf{2020}, \emph{12}, 50610--50618\relax
\mciteBstWouldAddEndPuncttrue
\mciteSetBstMidEndSepPunct{\mcitedefaultmidpunct}
{\mcitedefaultendpunct}{\mcitedefaultseppunct}\relax
\EndOfBibitem
\bibitem[Awais \latin{et~al.}(2020)Awais, Khan, Hassan, Bae, and
  Chattha]{Awais2020}
Awais,~M.; Khan,~M.~U.; Hassan,~A.; Bae,~J.; Chattha,~T.~E. {Printable Highly
  Stable and Superfast Humidity Sensor Based on Two Dimensional Molybdenum
  Diselenide}. \emph{Scientific Reports} \textbf{2020}, \emph{10}, 1--13\relax
\mciteBstWouldAddEndPuncttrue
\mciteSetBstMidEndSepPunct{\mcitedefaultmidpunct}
{\mcitedefaultendpunct}{\mcitedefaultseppunct}\relax
\EndOfBibitem
\bibitem[Bertolazzi \latin{et~al.}(2013)Bertolazzi, Krasnozhon, and
  Kis]{Bertolazzi2013}
Bertolazzi,~S.; Krasnozhon,~D.; Kis,~A. {Nonvolatile memory cells based on
  MoS$_2$/graphene heterostructures}. \emph{ACS Nano} \textbf{2013}, \emph{7},
  3246--3252\relax
\mciteBstWouldAddEndPuncttrue
\mciteSetBstMidEndSepPunct{\mcitedefaultmidpunct}
{\mcitedefaultendpunct}{\mcitedefaultseppunct}\relax
\EndOfBibitem
\bibitem[Yin \latin{et~al.}(2013)Yin, Zeng, Liu, He, Chen, and Zhang]{Yin2013}
Yin,~Z.; Zeng,~Z.; Liu,~J.; He,~Q.; Chen,~P.; Zhang,~H. {Memory devices using a
  mixture of MoS$_2$ and graphene oxide as the active layer}. \emph{Small}
  \textbf{2013}, \emph{9}, 727--731\relax
\mciteBstWouldAddEndPuncttrue
\mciteSetBstMidEndSepPunct{\mcitedefaultmidpunct}
{\mcitedefaultendpunct}{\mcitedefaultseppunct}\relax
\EndOfBibitem
\bibitem[Yoshida \latin{et~al.}(2015)Yoshida, Suzuki, Zhang, Nakano, and
  Iwasa]{Yoshida2015}
Yoshida,~M.; Suzuki,~R.; Zhang,~Y.; Nakano,~M.; Iwasa,~Y. {Memristive phase
  switching in two-dimensional 1T-TaS$_2$ crystals}. \emph{Science Advances}
  \textbf{2015}, \emph{1}, e1500606\relax
\mciteBstWouldAddEndPuncttrue
\mciteSetBstMidEndSepPunct{\mcitedefaultmidpunct}
{\mcitedefaultendpunct}{\mcitedefaultseppunct}\relax
\EndOfBibitem
\bibitem[Radisavljevic \latin{et~al.}(2011)Radisavljevic, Radenovic, Brivio,
  Giacometti, and Kis]{Radisavljevic2011}
Radisavljevic,~B.; Radenovic,~A.; Brivio,~J.; Giacometti,~V.; Kis,~A.
  {Single-layer MoS$_2$ transistors}. \emph{Nature Nanotechnology}
  \textbf{2011}, \emph{6}, 147--150\relax
\mciteBstWouldAddEndPuncttrue
\mciteSetBstMidEndSepPunct{\mcitedefaultmidpunct}
{\mcitedefaultendpunct}{\mcitedefaultseppunct}\relax
\EndOfBibitem
\bibitem[Yin \latin{et~al.}(2012)Yin, Li, Li, Jiang, Shi, Sun, Lu, Zhang, Chen,
  and Zhang]{Yin2012}
Yin,~Z.; Li,~H.; Li,~H.; Jiang,~L.; Shi,~Y.; Sun,~Y.; Lu,~G.; Zhang,~Q.;
  Chen,~X.; Zhang,~H. {Single-layer MoS$_2$ phototransistors}. \emph{ACS Nano}
  \textbf{2012}, \emph{6}, 74--80\relax
\mciteBstWouldAddEndPuncttrue
\mciteSetBstMidEndSepPunct{\mcitedefaultmidpunct}
{\mcitedefaultendpunct}{\mcitedefaultseppunct}\relax
\EndOfBibitem
\bibitem[Zhang \latin{et~al.}(2012)Zhang, Ye, Matsuhashi, and Iwasa]{Zhang2012}
Zhang,~Y.; Ye,~J.; Matsuhashi,~Y.; Iwasa,~Y. {Ambipolar MoS$_2$ thin flake
  transistors}. \emph{Nano Letters} \textbf{2012}, \emph{12}, 1136--1140\relax
\mciteBstWouldAddEndPuncttrue
\mciteSetBstMidEndSepPunct{\mcitedefaultmidpunct}
{\mcitedefaultendpunct}{\mcitedefaultseppunct}\relax
\EndOfBibitem
\bibitem[Novoselov \latin{et~al.}(2005)Novoselov, Jiang, Schedin, Booth,
  Khotkevich, Morozov, and Geim]{Novoselov2005}
Novoselov,~K.~S.; Jiang,~D.; Schedin,~F.; Booth,~T.~J.; Khotkevich,~V.~V.;
  Morozov,~S.~V.; Geim,~A.~K. {Two-dimensional atomic crystals}.
  \emph{Proceedings of the National Academy of Sciences of the United States of
  America} \textbf{2005}, \emph{102}, 10451--10453\relax
\mciteBstWouldAddEndPuncttrue
\mciteSetBstMidEndSepPunct{\mcitedefaultmidpunct}
{\mcitedefaultendpunct}{\mcitedefaultseppunct}\relax
\EndOfBibitem
\bibitem[Li \latin{et~al.}(2013)Li, Lu, Wang, Yin, Cong, He, Wang, Ding, Yu,
  and Zhang]{Li2013}
Li,~H.; Lu,~G.; Wang,~Y.; Yin,~Z.; Cong,~C.; He,~Q.; Wang,~L.; Ding,~F.;
  Yu,~T.; Zhang,~H. {Mechanical exfoliation and characterization of single- and
  few-layer nanosheets of WSe$_2$, TaS$_2$, and TaSe$_2$}. \emph{Small}
  \textbf{2013}, \emph{9}, 1974--1981\relax
\mciteBstWouldAddEndPuncttrue
\mciteSetBstMidEndSepPunct{\mcitedefaultmidpunct}
{\mcitedefaultendpunct}{\mcitedefaultseppunct}\relax
\EndOfBibitem
\bibitem[Li \latin{et~al.}(2014)Li, Wu, Yin, and Zhang]{Li2014}
Li,~H.; Wu,~J.; Yin,~Z.; Zhang,~H. {Preparation and applications of
  mechanically exfoliated single-layer and multilayer MoS$_2$ and WSe$_2$
  nanosheets}. \emph{Accounts of Chemical Research} \textbf{2014}, \emph{47},
  1067--1075\relax
\mciteBstWouldAddEndPuncttrue
\mciteSetBstMidEndSepPunct{\mcitedefaultmidpunct}
{\mcitedefaultendpunct}{\mcitedefaultseppunct}\relax
\EndOfBibitem
\bibitem[Yi and Shen(2015)Yi, and Shen]{Yi2015}
Yi,~M.; Shen,~Z. {A review on mechanical exfoliation for the scalable
  production of graphene}. \emph{Journal of Materials Chemistry A}
  \textbf{2015}, \emph{3}, 11700--11715\relax
\mciteBstWouldAddEndPuncttrue
\mciteSetBstMidEndSepPunct{\mcitedefaultmidpunct}
{\mcitedefaultendpunct}{\mcitedefaultseppunct}\relax
\EndOfBibitem
\bibitem[Nasiri and Tricoli(2019)Nasiri, and Tricoli]{Nasiri2019}
Nasiri,~N.; Tricoli,~A. \emph{Industrial Applications of Nanomaterials};
  Elsevier, 2019; pp 123--149\relax
\mciteBstWouldAddEndPuncttrue
\mciteSetBstMidEndSepPunct{\mcitedefaultmidpunct}
{\mcitedefaultendpunct}{\mcitedefaultseppunct}\relax
\EndOfBibitem
\bibitem[Cadiz \latin{et~al.}(2017)Cadiz, Courtade, Robert, Wang, Shen, Cai,
  Taniguchi, Watanabe, Carrere, Lagarde, Manca, Amand, Renucci, Tongay, Marie,
  and Urbaszek]{Cadiz}
Cadiz,~F. \latin{et~al.}  {Excitonic linewidth approaching the homogeneous
  limit in MoS$_2$-based van der Waals heterostructures}. \emph{Physical Review
  X} \textbf{2017}, \emph{7}\relax
\mciteBstWouldAddEndPuncttrue
\mciteSetBstMidEndSepPunct{\mcitedefaultmidpunct}
{\mcitedefaultendpunct}{\mcitedefaultseppunct}\relax
\EndOfBibitem
\bibitem[Courtade \latin{et~al.}(2018)Courtade, Han, Nakhaie, Robert, Marie,
  Renucci, Taniguchi, Watanabe, Geelhaar, Lopes, and Urbaszek]{Courtade2018}
Courtade,~E.; Han,~B.; Nakhaie,~S.; Robert,~C.; Marie,~X.; Renucci,~P.;
  Taniguchi,~T.; Watanabe,~K.; Geelhaar,~L.; Lopes,~J.~M.; Urbaszek,~B.
  {Spectrally narrow exciton luminescence from monolayer MoS$_2$ and MoSe$_2$
  exfoliated onto epitaxially grown hexagonal BN}. \emph{Applied Physics
  Letters} \textbf{2018}, \emph{113}, 32106\relax
\mciteBstWouldAddEndPuncttrue
\mciteSetBstMidEndSepPunct{\mcitedefaultmidpunct}
{\mcitedefaultendpunct}{\mcitedefaultseppunct}\relax
\EndOfBibitem
\bibitem[Grzeszczyk \latin{et~al.}(2020)Grzeszczyk, Molas, Barto{\v{s}},
  Nogajewski, Potemski, and Babi{\'{n}}ski]{Grzeszczyk2019}
Grzeszczyk,~M.; Molas,~M.~R.; Barto{\v{s}},~M.; Nogajewski,~K.; Potemski,~M.;
  Babi{\'{n}}ski,~A. {Breathing modes in few-layer MoTe$_2$ activated by h-BN
  encapsulation}. \emph{Applied Physics Letters} \textbf{2020}, \emph{116},
  191601\relax
\mciteBstWouldAddEndPuncttrue
\mciteSetBstMidEndSepPunct{\mcitedefaultmidpunct}
{\mcitedefaultendpunct}{\mcitedefaultseppunct}\relax
\EndOfBibitem
\bibitem[Han \latin{et~al.}(2019)Han, Lin, Liu, Wang, and Pan]{Han2019}
Han,~X.; Lin,~J.; Liu,~J.; Wang,~N.; Pan,~D. {Effects of Hexagonal Boron
  Nitride Encapsulation on the Electronic Structure of Few-Layer MoS$_2$}.
  \emph{Journal of Physical Chemistry C} \textbf{2019}, \emph{123},
  14797--14802\relax
\mciteBstWouldAddEndPuncttrue
\mciteSetBstMidEndSepPunct{\mcitedefaultmidpunct}
{\mcitedefaultendpunct}{\mcitedefaultseppunct}\relax
\EndOfBibitem
\bibitem[Hayashida \latin{et~al.}(2020)Hayashida, Saitoh, Watanabe, Taniguchi,
  Sawano, and Hoshi]{Hayashida2020}
Hayashida,~S.; Saitoh,~R.; Watanabe,~K.; Taniguchi,~T.; Sawano,~K.; Hoshi,~Y.
  {Reduced Inhomogeneous Broadening in Hexagonal Boron Nitride-Encapsulated
  MoTe$_2$ Monolayers by Thermal Treatment}. \emph{ACS Applied Electronic
  Materials} \textbf{2020}, \emph{2}, 2739--2744\relax
\mciteBstWouldAddEndPuncttrue
\mciteSetBstMidEndSepPunct{\mcitedefaultmidpunct}
{\mcitedefaultendpunct}{\mcitedefaultseppunct}\relax
\EndOfBibitem
\bibitem[Wang \latin{et~al.}(2020)Wang, Li, Ko, Shawkat, Okogbue, Yoo, Han,
  Islam, Oh, and Jung]{Wang2020}
Wang,~M.; Li,~H.; Ko,~T.~J.; Shawkat,~M.~S.; Okogbue,~E.; Yoo,~C.; Han,~S.~S.;
  Islam,~M.~A.; Oh,~K.~H.; Jung,~Y. {Manufacturing strategies for wafer-scale
  two-dimensional transition metal dichalcogenide heterolayers}. \emph{Journal
  of Materials Research} \textbf{2020}, \emph{35}, 1350--1368\relax
\mciteBstWouldAddEndPuncttrue
\mciteSetBstMidEndSepPunct{\mcitedefaultmidpunct}
{\mcitedefaultendpunct}{\mcitedefaultseppunct}\relax
\EndOfBibitem
\bibitem[Mandyam \latin{et~al.}(2020)Mandyam, Kim, and
  Drndi{\'{c}}]{Mandyam2020}
Mandyam,~S.~V.; Kim,~H.~M.; Drndi{\'{c}},~M. {Large area few-layer TMD film
  growths and their applications}. \emph{Journal of Physics: Materials}
  \textbf{2020}, \emph{3}, 024008\relax
\mciteBstWouldAddEndPuncttrue
\mciteSetBstMidEndSepPunct{\mcitedefaultmidpunct}
{\mcitedefaultendpunct}{\mcitedefaultseppunct}\relax
\EndOfBibitem
\bibitem[Zhan \latin{et~al.}(2012)Zhan, Liu, Najmaei, Ajayan, and
  Lou]{Zhan2012}
Zhan,~Y.; Liu,~Z.; Najmaei,~S.; Ajayan,~P.~M.; Lou,~J. {Large-area vapor-phase
  growth and characterization of MoS$_2$ atomic layers on a SiO$_2$ substrate}.
  \emph{Small} \textbf{2012}, \emph{8}, 966--971\relax
\mciteBstWouldAddEndPuncttrue
\mciteSetBstMidEndSepPunct{\mcitedefaultmidpunct}
{\mcitedefaultendpunct}{\mcitedefaultseppunct}\relax
\EndOfBibitem
\bibitem[Duan \latin{et~al.}(2014)Duan, Wang, Shaw, Cheng, Chen, Li, Wu, Tang,
  Zhang, Pan, Jiang, Yu, Huang, and Duan]{Duan2014}
Duan,~X.; Wang,~C.; Shaw,~J.~C.; Cheng,~R.; Chen,~Y.; Li,~H.; Wu,~X.; Tang,~Y.;
  Zhang,~Q.; Pan,~A.; Jiang,~J.; Yu,~R.; Huang,~Y.; Duan,~X. {Lateral epitaxial
  growth of two-dimensional layered semiconductor heterojunctions}.
  \emph{Nature Nanotechnology} \textbf{2014}, \emph{9}, 1024--1030\relax
\mciteBstWouldAddEndPuncttrue
\mciteSetBstMidEndSepPunct{\mcitedefaultmidpunct}
{\mcitedefaultendpunct}{\mcitedefaultseppunct}\relax
\EndOfBibitem
\bibitem[Liu \latin{et~al.}(2015)Liu, Wong, and Chi]{Liu2015}
Liu,~H.~F.; Wong,~S.~L.; Chi,~D.~Z. {CVD Growth of MoS$_2$-based
  Two-dimensional Materials}. \emph{Chemical Vapor Deposition} \textbf{2015},
  \emph{21}, 241--259\relax
\mciteBstWouldAddEndPuncttrue
\mciteSetBstMidEndSepPunct{\mcitedefaultmidpunct}
{\mcitedefaultendpunct}{\mcitedefaultseppunct}\relax
\EndOfBibitem
\bibitem[Jeon \latin{et~al.}(2015)Jeon, Jang, Jeon, Yoo, Jang, Park, and
  Lee]{Jeon2015}
Jeon,~J.; Jang,~S.~K.; Jeon,~S.~M.; Yoo,~G.; Jang,~Y.~H.; Park,~J.~H.; Lee,~S.
  {Layer-controlled CVD growth of large-area two-dimensional MoS$_2$ films}.
  \emph{Nanoscale} \textbf{2015}, \emph{7}, 1688--1695\relax
\mciteBstWouldAddEndPuncttrue
\mciteSetBstMidEndSepPunct{\mcitedefaultmidpunct}
{\mcitedefaultendpunct}{\mcitedefaultseppunct}\relax
\EndOfBibitem
\bibitem[Yu \latin{et~al.}(2017)Yu, Liao, Zhao, Liu, Zhou, Wei, Xu, Liu, Hu,
  Deng, Zhou, Shi, Gu, Shen, Zhang, Du, Xie, Zhu, Chen, Yang, Shi, and
  Zhang]{Yu2017}
Yu,~H. \latin{et~al.}  {Wafer-Scale Growth and Transfer of Highly-Oriented
  Monolayer MoS$_2$ Continuous Films}. \emph{ACS Nano} \textbf{2017},
  \emph{11}, 12001--12007\relax
\mciteBstWouldAddEndPuncttrue
\mciteSetBstMidEndSepPunct{\mcitedefaultmidpunct}
{\mcitedefaultendpunct}{\mcitedefaultseppunct}\relax
\EndOfBibitem
\bibitem[Wang \latin{et~al.}(2020)Wang, Li, Tang, Zhu, Zhang, Jia, Lu, Wei, Yu,
  Zhao, Guo, Gu, Sun, Yang, Yang, Shi, and Zhang]{Wang2020a}
Wang,~Q. \latin{et~al.}  {Wafer-Scale Highly Oriented Monolayer MoS$_2$ with
  Large Domain Sizes}. \emph{Nano Letters} \textbf{2020}, \emph{20},
  7193--7199\relax
\mciteBstWouldAddEndPuncttrue
\mciteSetBstMidEndSepPunct{\mcitedefaultmidpunct}
{\mcitedefaultendpunct}{\mcitedefaultseppunct}\relax
\EndOfBibitem
\bibitem[Delhomme \latin{et~al.}(2019)Delhomme, Butseraen, Zheng, Marty,
  Bouchiat, Molas, Pan, Watanabe, Taniguchi, Ouerghi, Renard, and
  Faugeras]{Delhomme2019}
Delhomme,~A.; Butseraen,~G.; Zheng,~B.; Marty,~L.; Bouchiat,~V.; Molas,~M.~R.;
  Pan,~A.; Watanabe,~K.; Taniguchi,~T.; Ouerghi,~A.; Renard,~J.; Faugeras,~C.
  {Magneto-spectroscopy of exciton Rydberg states in a CVD grown WSe$_2$
  monolayer}. \emph{Applied Physics Letters} \textbf{2019}, \emph{114},
  232104\relax
\mciteBstWouldAddEndPuncttrue
\mciteSetBstMidEndSepPunct{\mcitedefaultmidpunct}
{\mcitedefaultendpunct}{\mcitedefaultseppunct}\relax
\EndOfBibitem
\bibitem[Shree \latin{et~al.}(2020)Shree, George, Lehnert, Neumann, Benelajla,
  Robert, Marie, Watanabe, Taniguchi, Kaiser, Urbaszek, and
  Turchanin]{Shree2020}
Shree,~S.; George,~A.; Lehnert,~T.; Neumann,~C.; Benelajla,~M.; Robert,~C.;
  Marie,~X.; Watanabe,~K.; Taniguchi,~T.; Kaiser,~U.; Urbaszek,~B.;
  Turchanin,~A. {High optical quality of MoS$_2$ monolayers grown by chemical
  vapor deposition}. \emph{2D Materials} \textbf{2020}, \emph{7}, 015011\relax
\mciteBstWouldAddEndPuncttrue
\mciteSetBstMidEndSepPunct{\mcitedefaultmidpunct}
{\mcitedefaultendpunct}{\mcitedefaultseppunct}\relax
\EndOfBibitem
\bibitem[Marx \latin{et~al.}(2017)Marx, Nordmann, Knoch, Franzen, Stampfer,
  Andrzejewski, K{\"{u}}mmell, Bacher, Heuken, Kalisch, and Vescan]{Marx2017}
Marx,~M.; Nordmann,~S.; Knoch,~J.; Franzen,~C.; Stampfer,~C.; Andrzejewski,~D.;
  K{\"{u}}mmell,~T.; Bacher,~G.; Heuken,~M.; Kalisch,~H.; Vescan,~A.
  {Large-area MoS$_2$ deposition via MOVPE}. \emph{Journal of Crystal Growth}
  \textbf{2017}, \emph{464}, 100--104\relax
\mciteBstWouldAddEndPuncttrue
\mciteSetBstMidEndSepPunct{\mcitedefaultmidpunct}
{\mcitedefaultendpunct}{\mcitedefaultseppunct}\relax
\EndOfBibitem
\bibitem[Xenogiannopoulou \latin{et~al.}(2015)Xenogiannopoulou, Tsipas,
  Aretouli, Tsoutsou, Giamini, Bazioti, Dimitrakopulos, Komninou, Brems,
  Huyghebaert, Radu, and Dimoulas]{Xenogiannopoulou2015}
Xenogiannopoulou,~E.; Tsipas,~P.; Aretouli,~K.~E.; Tsoutsou,~D.;
  Giamini,~S.~A.; Bazioti,~C.; Dimitrakopulos,~G.~P.; Komninou,~P.; Brems,~S.;
  Huyghebaert,~C.; Radu,~I.~P.; Dimoulas,~A. {High-quality, large-area MoSe$_2$
  and MoSe$_2$/Bi2Se$_3$ heterostructures on AlN(0001)/Si(111) substrates by
  molecular beam epitaxy}. \emph{Nanoscale} \textbf{2015}, \emph{7},
  7896--7905\relax
\mciteBstWouldAddEndPuncttrue
\mciteSetBstMidEndSepPunct{\mcitedefaultmidpunct}
{\mcitedefaultendpunct}{\mcitedefaultseppunct}\relax
\EndOfBibitem
\bibitem[Ehlen \latin{et~al.}(2019)Ehlen, Hall, Senkovskiy, Hell, Li, Herman,
  Smirnov, Fedorov, {Yu Voroshnin}, {Di Santo}, Petaccia, Michely, and
  Gr{\"{u}}neis]{Ehlen2019}
Ehlen,~N.; Hall,~J.; Senkovskiy,~B.~V.; Hell,~M.; Li,~J.; Herman,~A.;
  Smirnov,~D.; Fedorov,~A.; {Yu Voroshnin},~V.; {Di Santo},~G.; Petaccia,~L.;
  Michely,~T.; Gr{\"{u}}neis,~A. {Narrow photoluminescence and Raman peaks of
  epitaxial MoS$_2$ on graphene/Ir(1 1 1)}. \emph{2D Materials} \textbf{2019},
  \emph{6}, 011006\relax
\mciteBstWouldAddEndPuncttrue
\mciteSetBstMidEndSepPunct{\mcitedefaultmidpunct}
{\mcitedefaultendpunct}{\mcitedefaultseppunct}\relax
\EndOfBibitem
\bibitem[Wei \latin{et~al.}(2020)Wei, Hu, Li, Hu, Yu, Sun, Hohage, and
  Sun]{Wei2020}
Wei,~Y.; Hu,~C.; Li,~Y.; Hu,~X.; Yu,~K.; Sun,~L.; Hohage,~M.; Sun,~L. {Initial
  stage of MBE growth of MoSe$_2$ monolayer}. \emph{Nanotechnology}
  \textbf{2020}, \emph{31}, 315710\relax
\mciteBstWouldAddEndPuncttrue
\mciteSetBstMidEndSepPunct{\mcitedefaultmidpunct}
{\mcitedefaultendpunct}{\mcitedefaultseppunct}\relax
\EndOfBibitem
\bibitem[Wei \latin{et~al.}(2020)Wei, Hu, Li, Hu, Hohage, and Sun]{Wei2020a}
Wei,~Y.; Hu,~C.; Li,~Y.; Hu,~X.; Hohage,~M.; Sun,~L. {Growth oscillation of
  MoSe$_2$ monolayers observed by differential reflectance spectroscopy}.
  \emph{Journal of Physics Condensed Matter} \textbf{2020}, \emph{32},
  155001\relax
\mciteBstWouldAddEndPuncttrue
\mciteSetBstMidEndSepPunct{\mcitedefaultmidpunct}
{\mcitedefaultendpunct}{\mcitedefaultseppunct}\relax
\EndOfBibitem
\bibitem[Ohtake and Sakuma(2020)Ohtake, and Sakuma]{Ohtake2020}
Ohtake,~A.; Sakuma,~Y. {Effect of Substrate Orientation on MoSe$_2$/GaAs
  Heteroepitaxy}. \emph{Journal of Physical Chemistry C} \textbf{2020},
  \emph{124}, 5203\relax
\mciteBstWouldAddEndPuncttrue
\mciteSetBstMidEndSepPunct{\mcitedefaultmidpunct}
{\mcitedefaultendpunct}{\mcitedefaultseppunct}\relax
\EndOfBibitem
\bibitem[He \latin{et~al.}(2020)He, Wei, Huang, Zhou, Hu, Xie, Chen, Wu, and
  Li]{He2020}
He,~Z.; Wei,~T.; Huang,~W.; Zhou,~W.; Hu,~P.; Xie,~Z.; Chen,~H.; Wu,~S.; Li,~S.
  {Electrostatically Enhanced Electron-Phonon Interaction in Monolayer
  2H-MoSe$_2$ Grown by Molecular Beam Epitaxy}. \emph{ACS Applied Materials and
  Interfaces} \textbf{2020}, \emph{12}, 44067--44073\relax
\mciteBstWouldAddEndPuncttrue
\mciteSetBstMidEndSepPunct{\mcitedefaultmidpunct}
{\mcitedefaultendpunct}{\mcitedefaultseppunct}\relax
\EndOfBibitem
\bibitem[Ogorzalek \latin{et~al.}(2020)Ogorzalek, Seredynski, Kret,
  Kwiatkowski, Korona, Grzeszczyk, Mierzejewski, Wasik, Pacuski, Sadowski, and
  Gryglas-Borysiewicz]{Ogorzalek2020}
Ogorzalek,~Z.; Seredynski,~B.; Kret,~S.; Kwiatkowski,~A.; Korona,~K.~P.;
  Grzeszczyk,~M.; Mierzejewski,~J.; Wasik,~D.; Pacuski,~W.; Sadowski,~J.;
  Gryglas-Borysiewicz,~M. {Charge transport in MBE-grown 2H-MoTe$_2$ bilayers
  with enhanced stability provided by an AlO: Xcapping layer}. \emph{Nanoscale}
  \textbf{2020}, \emph{12}, 16535--16542\relax
\mciteBstWouldAddEndPuncttrue
\mciteSetBstMidEndSepPunct{\mcitedefaultmidpunct}
{\mcitedefaultendpunct}{\mcitedefaultseppunct}\relax
\EndOfBibitem
\bibitem[Yan \latin{et~al.}(2015)Yan, Velasco, Kahn, Watanabe, Taniguchi, Wang,
  Crommie, and Zettl]{Yan2015}
Yan,~A.; Velasco,~J.; Kahn,~S.; Watanabe,~K.; Taniguchi,~T.; Wang,~F.;
  Crommie,~M.~F.; Zettl,~A. {Direct Growth of Single- and Few-Layer MoS$_2$ on
  h-BN with Preferred Relative Rotation Angles}. \emph{Nano Letters}
  \textbf{2015}, \emph{15}, 6324--6331\relax
\mciteBstWouldAddEndPuncttrue
\mciteSetBstMidEndSepPunct{\mcitedefaultmidpunct}
{\mcitedefaultendpunct}{\mcitedefaultseppunct}\relax
\EndOfBibitem
\bibitem[Pacuski \latin{et~al.}(2020)Pacuski, Grzeszczyk, Nogajewski, Bogucki,
  Oreszczuk, Kucharek, Po{\l}czy{\'{n}}ska, Seredy{\'{n}}ski, Rodek,
  Bo{\.{z}}ek, Taniguchi, Watanabe, Kret, Sadowski, Kazimierczuk, Potemski, and
  Kossacki]{Pacuski2020}
Pacuski,~W. \latin{et~al.}  {Narrow Excitonic Lines and Large-Scale Homogeneity
  of Transition-Metal Dichalcogenide Monolayers Grown by Molecular Beam Epitaxy
  on Hexagonal Boron Nitride}. \emph{Nano Letters} \textbf{2020}, \emph{20},
  3058--3066\relax
\mciteBstWouldAddEndPuncttrue
\mciteSetBstMidEndSepPunct{\mcitedefaultmidpunct}
{\mcitedefaultendpunct}{\mcitedefaultseppunct}\relax
\EndOfBibitem
\bibitem[Fu \latin{et~al.}(2017)Fu, Zhao, Zhang, Li, Xu, Jang, Yoon, Song, Poh,
  Ren, Ding, Fu, Shin, Shin, Pantelides, Zhou, and Loh]{Fu2017}
Fu,~D. \latin{et~al.}  {Molecular Beam Epitaxy of Highly Crystalline Monolayer
  Molybdenum Disulfide on Hexagonal Boron Nitride}. \emph{Journal of the
  American Chemical Society} \textbf{2017}, \emph{139}, 9392--9400\relax
\mciteBstWouldAddEndPuncttrue
\mciteSetBstMidEndSepPunct{\mcitedefaultmidpunct}
{\mcitedefaultendpunct}{\mcitedefaultseppunct}\relax
\EndOfBibitem
\bibitem[D{\c{a}}browska \latin{et~al.}(2020)D{\c{a}}browska, Tokarczyk,
  Kowalski, Binder, Bozek, Borysiuk, Stępniewski, and
  Wysmo{\l}ek]{Dabrowska2020}
D{\c{a}}browska,~A.~K.; Tokarczyk,~M.; Kowalski,~G.; Binder,~J.; Bozek,~R.;
  Borysiuk,~J.; Stępniewski,~R.; Wysmo{\l}ek,~A. {Two stage epitaxial growth
  of wafer-size multilayer h-BN by metal-organic vapor phase epitaxy - A
  homoepitaxial approach}. \emph{2D Materials} \textbf{2020}, \emph{8},
  15017\relax
\mciteBstWouldAddEndPuncttrue
\mciteSetBstMidEndSepPunct{\mcitedefaultmidpunct}
{\mcitedefaultendpunct}{\mcitedefaultseppunct}\relax
\EndOfBibitem
\bibitem[Mak \latin{et~al.}(2010)Mak, Lee, Hone, Shan, and Heinz]{Mak2010}
Mak,~K.~F.; Lee,~C.; Hone,~J.; Shan,~J.; Heinz,~T.~F. {Atomically thin MoS$_2$:
  A new direct-gap semiconductor}. \emph{Physical Review Letters}
  \textbf{2010}, \emph{105}\relax
\mciteBstWouldAddEndPuncttrue
\mciteSetBstMidEndSepPunct{\mcitedefaultmidpunct}
{\mcitedefaultendpunct}{\mcitedefaultseppunct}\relax
\EndOfBibitem
\bibitem[Splendiani \latin{et~al.}(2010)Splendiani, Sun, Zhang, Li, Kim, Chim,
  Galli, and Wang]{Splendiani2010}
Splendiani,~A.; Sun,~L.; Zhang,~Y.; Li,~T.; Kim,~J.; Chim,~C.~Y.; Galli,~G.;
  Wang,~F. {Emerging photoluminescence in monolayer MoS$_2$}. \emph{Nano
  Letters} \textbf{2010}, \emph{10}, 1271--1275\relax
\mciteBstWouldAddEndPuncttrue
\mciteSetBstMidEndSepPunct{\mcitedefaultmidpunct}
{\mcitedefaultendpunct}{\mcitedefaultseppunct}\relax
\EndOfBibitem
\bibitem[Tonndorf \latin{et~al.}(2013)Tonndorf, Schmidt, B{\"{o}}ttger, Zhang,
  B{\"{o}}rner, Liebig, Albrecht, Kloc, Gordan, Zahn, {Michaelis de
  Vasconcellos}, and Bratschitsch]{Tonndorf2013}
Tonndorf,~P.; Schmidt,~R.; B{\"{o}}ttger,~P.; Zhang,~X.; B{\"{o}}rner,~J.;
  Liebig,~A.; Albrecht,~M.; Kloc,~C.; Gordan,~O.; Zahn,~D. R.~T.; {Michaelis de
  Vasconcellos},~S.; Bratschitsch,~R. {Photoluminescence emission and Raman
  response of monolayer MoS$_2$, MoSe$_2$, and WSe$_2$}. \emph{Optics Express}
  \textbf{2013}, \emph{21}, 4908\relax
\mciteBstWouldAddEndPuncttrue
\mciteSetBstMidEndSepPunct{\mcitedefaultmidpunct}
{\mcitedefaultendpunct}{\mcitedefaultseppunct}\relax
\EndOfBibitem
\bibitem[Zhang \latin{et~al.}(2016)Zhang, Chen, Zhang, Pan, Chou, Zeng, and
  Shih]{Zhang}
Zhang,~Q.; Chen,~Y.; Zhang,~C.; Pan,~C.~R.; Chou,~M.~Y.; Zeng,~C.; Shih,~C.~K.
  {Bandgap renormalization and work function tuning in MoSe$_2$/hBN/Ru(0001)
  heterostructures}. \emph{Nature Communications} \textbf{2016}, \emph{7}\relax
\mciteBstWouldAddEndPuncttrue
\mciteSetBstMidEndSepPunct{\mcitedefaultmidpunct}
{\mcitedefaultendpunct}{\mcitedefaultseppunct}\relax
\EndOfBibitem
\bibitem[Cadiz \latin{et~al.}(2016)Cadiz, Robert, Wang, Kong, Fan, Blei,
  Lagarde, Gay, Manca, Taniguchi, Watanabe, Amand, Marie, Renucci, Tongay, and
  Urbaszek]{Cadiz2016}
Cadiz,~F. \latin{et~al.}  {Ultra-low power threshold for laser induced changes
  in optical properties of 2D molybdenum dichalcogenides}. \emph{2D Materials}
  \textbf{2016}, \emph{3}\relax
\mciteBstWouldAddEndPuncttrue
\mciteSetBstMidEndSepPunct{\mcitedefaultmidpunct}
{\mcitedefaultendpunct}{\mcitedefaultseppunct}\relax
\EndOfBibitem
\bibitem[Pakula \latin{et~al.}(2019)Pakula, Dabrowska, Tokarczyk, Bozek,
  Binder, Kowalski, Wysmolek, and Stepniewski]{Pakua2019}
Pakula,~K.; Dabrowska,~A.; Tokarczyk,~M.; Bozek,~R.; Binder,~J.; Kowalski,~G.;
  Wysmolek,~A.; Stepniewski,~R. {Fundamental mechanisms of hBN growth by
  MOVPE}. \emph{arXiv} \textbf{2019}, \relax
\mciteBstWouldAddEndPunctfalse
\mciteSetBstMidEndSepPunct{\mcitedefaultmidpunct}
{}{\mcitedefaultseppunct}\relax
\EndOfBibitem
\bibitem[Nagashima \latin{et~al.}(1995)Nagashima, Tejima, Gamou, Kawai, and
  Oshima]{Nagashima1995}
Nagashima,~A.; Tejima,~N.; Gamou,~Y.; Kawai,~T.; Oshima,~C. {Electronic
  dispersion relations of monolayer hexagonal boron nitride formed on the
  Ni(111) surface}. \emph{Physical Review B} \textbf{1995}, \emph{51},
  4606--4613\relax
\mciteBstWouldAddEndPuncttrue
\mciteSetBstMidEndSepPunct{\mcitedefaultmidpunct}
{\mcitedefaultendpunct}{\mcitedefaultseppunct}\relax
\EndOfBibitem
\bibitem[Geick \latin{et~al.}(1966)Geick, Perry, and Rupprecht]{Geick1966}
Geick,~R.; Perry,~C.~H.; Rupprecht,~G. {Normal modes in hexagonal boron
  nitride}. \emph{Physical Review} \textbf{1966}, \emph{146}, 543--547\relax
\mciteBstWouldAddEndPuncttrue
\mciteSetBstMidEndSepPunct{\mcitedefaultmidpunct}
{\mcitedefaultendpunct}{\mcitedefaultseppunct}\relax
\EndOfBibitem
\bibitem[Kobayashi and Akasaka(2008)Kobayashi, and Akasaka]{Kobayashi2008}
Kobayashi,~Y.; Akasaka,~T. {Hexagonal BN epitaxial growth on (0 0 0 1) sapphire
  substrate by MOVPE}. \emph{Journal of Crystal Growth} \textbf{2008},
  \emph{310}, 5044--5047\relax
\mciteBstWouldAddEndPuncttrue
\mciteSetBstMidEndSepPunct{\mcitedefaultmidpunct}
{\mcitedefaultendpunct}{\mcitedefaultseppunct}\relax
\EndOfBibitem
\bibitem[Ross \latin{et~al.}(2013)Ross, Wu, Yu, Ghimire, Jones, Aivazian, Yan,
  Mandrus, Xiao, Yao, and Xu]{Ross2013}
Ross,~J.~S.; Wu,~S.; Yu,~H.; Ghimire,~N.~J.; Jones,~A.~M.; Aivazian,~G.;
  Yan,~J.; Mandrus,~D.~G.; Xiao,~D.; Yao,~W.; Xu,~X. {Electrical control of
  neutral and charged excitons in a monolayer semiconductor}. \emph{Nature
  Communications} \textbf{2013}, \emph{4}\relax
\mciteBstWouldAddEndPuncttrue
\mciteSetBstMidEndSepPunct{\mcitedefaultmidpunct}
{\mcitedefaultendpunct}{\mcitedefaultseppunct}\relax
\EndOfBibitem
\bibitem[Zhang \latin{et~al.}(2018)Zhang, Sharma, Zhu, Zhang, Wang, Dong,
  Nguyen, Wang, Wen, Cao, Liu, Sun, Yang, Li, Kar, Shi, Macdonald, Yu, Wang,
  and Lu]{Zhang2018}
Zhang,~L. \latin{et~al.}  {Efficient and Layer-Dependent Exciton Pumping across
  Atomically Thin Organic–Inorganic Type-I Heterostructures}. \emph{Advanced
  Materials} \textbf{2018}, \emph{30}\relax
\mciteBstWouldAddEndPuncttrue
\mciteSetBstMidEndSepPunct{\mcitedefaultmidpunct}
{\mcitedefaultendpunct}{\mcitedefaultseppunct}\relax
\EndOfBibitem
\bibitem[Ye \latin{et~al.}(2018)Ye, Yan, Niu, Li, and Zhang]{Ye2018}
Ye,~J.; Yan,~T.; Niu,~B.; Li,~Y.; Zhang,~X. {Nonlinear dynamics of trions under
  strong optical excitation in monolayer MoSe$_2$}. \emph{Scientific Reports}
  \textbf{2018}, \emph{8}, 1--8\relax
\mciteBstWouldAddEndPuncttrue
\mciteSetBstMidEndSepPunct{\mcitedefaultmidpunct}
{\mcitedefaultendpunct}{\mcitedefaultseppunct}\relax
\EndOfBibitem
\bibitem[Macfarlane(1963)]{Macfarlane1963}
Macfarlane,~R.~M. {Analysis of the Spectrum of d3 Ions in Trigonal Crystal
  Fields}. \emph{The Journal of Chemical Physics} \textbf{1963}, \emph{39},
  3118\relax
\mciteBstWouldAddEndPuncttrue
\mciteSetBstMidEndSepPunct{\mcitedefaultmidpunct}
{\mcitedefaultendpunct}{\mcitedefaultseppunct}\relax
\EndOfBibitem
\bibitem[Wang \latin{et~al.}(2015)Wang, Palleau, Amand, Tongay, Marie, and
  Urbaszek]{Wang2015}
Wang,~G.; Palleau,~E.; Amand,~T.; Tongay,~S.; Marie,~X.; Urbaszek,~B.
  {Polarization and time-resolved photoluminescence spectroscopy of excitons in
  MoSe$_2$ monolayers}. \emph{Applied Physics Letters} \textbf{2015},
  \emph{106}, 112101\relax
\mciteBstWouldAddEndPuncttrue
\mciteSetBstMidEndSepPunct{\mcitedefaultmidpunct}
{\mcitedefaultendpunct}{\mcitedefaultseppunct}\relax
\EndOfBibitem
\bibitem[Wierzbowski \latin{et~al.}(2017)Wierzbowski, Klein, Sigger,
  Straubinger, Kremser, Taniguchi, Watanabe, Wurstbauer, Holleitner, Kaniber,
  M{\"{u}}ller, and Finley]{Wierzbowski2017}
Wierzbowski,~J.; Klein,~J.; Sigger,~F.; Straubinger,~C.; Kremser,~M.;
  Taniguchi,~T.; Watanabe,~K.; Wurstbauer,~U.; Holleitner,~A.~W.; Kaniber,~M.;
  M{\"{u}}ller,~K.; Finley,~J.~J. {Direct exciton emission from atomically thin
  transition metal dichalcogenide heterostructures near the lifetime limit}.
  \emph{Scientific Reports} \textbf{2017}, \emph{7}, 1--6\relax
\mciteBstWouldAddEndPuncttrue
\mciteSetBstMidEndSepPunct{\mcitedefaultmidpunct}
{\mcitedefaultendpunct}{\mcitedefaultseppunct}\relax
\EndOfBibitem
\bibitem[Godde \latin{et~al.}(2016)Godde, Schmidt, Schmutzler, A{\ss}mann,
  Debus, Withers, Alexeev, {Del Pozo-Zamudio}, Skrypka, Novoselov, Bayer, and
  Tartakovskii]{Godde2016}
Godde,~T.; Schmidt,~D.; Schmutzler,~J.; A{\ss}mann,~M.; Debus,~J.; Withers,~F.;
  Alexeev,~E.~M.; {Del Pozo-Zamudio},~O.; Skrypka,~O.~V.; Novoselov,~K.~S.;
  Bayer,~M.; Tartakovskii,~A.~I. {Exciton and trion dynamics in atomically thin
  MoSe$_2$ and WSe$_2$: Effect of localization}. \emph{Physical Review B}
  \textbf{2016}, \emph{94}, 165301\relax
\mciteBstWouldAddEndPuncttrue
\mciteSetBstMidEndSepPunct{\mcitedefaultmidpunct}
{\mcitedefaultendpunct}{\mcitedefaultseppunct}\relax
\EndOfBibitem
\bibitem[Zhang and Li(2019)Zhang, and Li]{Zhang2019}
Zhang,~X.; Li,~X.~B. {Characterization of Layer Number of Two-Dimensional
  Transition Metal Diselenide Semiconducting Devices Using Si-Peak Analysis}.
  \emph{Advances in Materials Science and Engineering} \textbf{2019},
  \emph{2019}\relax
\mciteBstWouldAddEndPuncttrue
\mciteSetBstMidEndSepPunct{\mcitedefaultmidpunct}
{\mcitedefaultendpunct}{\mcitedefaultseppunct}\relax
\EndOfBibitem
\bibitem[Matte \latin{et~al.}(2011)Matte, Plowman, Datta, and Rao]{Matte2011}
Matte,~H.~S.; Plowman,~B.; Datta,~R.; Rao,~C.~N. {Graphene analogues of layered
  metal selenides}. \emph{Dalton Transactions} \textbf{2011}, \emph{40},
  10322--10325\relax
\mciteBstWouldAddEndPuncttrue
\mciteSetBstMidEndSepPunct{\mcitedefaultmidpunct}
{\mcitedefaultendpunct}{\mcitedefaultseppunct}\relax
\EndOfBibitem
\bibitem[Loh \latin{et~al.}(2018)Loh, Poh, Zhao, {Rong Tan}, Fu, Fei, Chu,
  Jiadong, Zhou, Pennycook, and {Castro Neto}]{MuiPoh2018}
Loh,~K.~P.; Poh,~S.~M.; Zhao,~X.; {Rong Tan},~S.~J.; Fu,~D.; Fei,~W.; Chu,~L.;
  Jiadong,~D.; Zhou,~W.; Pennycook,~S.~J.; {Castro Neto},~A.~H. {Molecular beam
  epitaxy of highly crystalline MoSe$_2$ on hexagonal boron nitride}. \emph{ACS
  Nano} \textbf{2018}, \emph{12}, 7562--7570\relax
\mciteBstWouldAddEndPuncttrue
\mciteSetBstMidEndSepPunct{\mcitedefaultmidpunct}
{\mcitedefaultendpunct}{\mcitedefaultseppunct}\relax
\EndOfBibitem
\bibitem[Yun \latin{et~al.}(2012)Yun, Han, Hong, Kim, and Lee]{Yun2012}
Yun,~W.~S.; Han,~S.~W.; Hong,~S.~C.; Kim,~I.~G.; Lee,~J.~D. {Thickness and
  strain effects on electronic structures of transition metal dichalcogenides:
  2H-MX$_2$ semiconductors (M = Mo, W; X = S, Se, Te)}. \emph{Physical Review B
  - Condensed Matter and Materials Physics} \textbf{2012}, \emph{85}\relax
\mciteBstWouldAddEndPuncttrue
\mciteSetBstMidEndSepPunct{\mcitedefaultmidpunct}
{\mcitedefaultendpunct}{\mcitedefaultseppunct}\relax
\EndOfBibitem
\end{mcitethebibliography}

\end{document}